\documentclass[twocolumn,aps,prl,superscriptaddress,tightlines,nofootinbib,nobalancelastpage]{revtex4-2}
\usepackage{hyperref}
\hypersetup{
colorlinks=true,
linkcolor=blue,         
citecolor=blue,         
filecolor=blue,      
urlcolor=blue
}
\usepackage{echodefs}
\usepackage{url}
\usepackage{float}
\usepackage{placeins}
\usepackage{color}
\usepackage{graphicx,epsfig,colordvi,epstopdf}
\usepackage{bm}
\usepackage{mathtools}
\usepackage{mathdots}
\usepackage{cleveref}
\usepackage{microtype}
\usepackage{xcolor}
\usepackage[T1]{fontenc}

\renewcommand{\vec}[1]{\ensuremath{\boldsymbol{#1}}}
\providecommand{\vtheta}{\vec{\theta}}

\newcommand{\mathbbm}[1]{\text{\usefont{U}{bbm}{m}{n}#1}}

\definecolor{revgreen}{RGB}{20,150,20}
\definecolor{revblue}{RGB}{30,30,140}
\newcommand{\new}[1]{{#1}}

\makeatletter
\def\maketitle{
	\@author@finish
	\title@column\titleblock@produce
	\suppressfloats[t]}
\makeatother

\begin{document}

\title{Spring--block theory of feature learning in deep neural networks}
\author{Cheng Shi}
\affiliation{Departement Mathematik und Informatik, University of Basel, Spiegelgasse 1, 4051 Basel, Switzerland}
\author{Liming Pan}
\affiliation{School of Cyber Science and Technology, University of Science and Technology of China, 230026, Hefei, China}
\author{Ivan Dokmani\'c}\
\email{ivan.dokmanic@unibas.ch}
\affiliation{Departement Mathematik und Informatik, University of Basel, Spiegelgasse 1, 4051 Basel, Switzerland}
\affiliation{Department of Electrical and Computer Engineering, University of Illinois at Urbana-Champaign, 306 N Wright St, Urbana, IL 61801, USA}

\date{\today}

\begin{abstract}
Feature-learning deep nets progressively collapse data to a regular low-dimensional geometry. How this emerges from the collective action of nonlinearity, noise, learning rate, and other factors, has eluded first-principles theories built from microscopic neuronal dynamics. We exhibit a noise–nonlinearity phase diagram that identifies regimes where shallow or deep layers learn more effectively and propose a macroscopic mechanical theory that reproduces the diagram and links feature learning across layers to generalization.
\end{abstract}

\maketitle

Deep neural networks (DNNs) progressively compute features from which the final layer generates predictions. When optimized via stochastic dynamics over a data-dependent energy, each layer learns to compute better features than the previous one~\cite{zeiler2014visualizing}, ultimately transforming the data to a regular low-dimensional geometry~\cite{papyan2020prevalence,he2023law,zarka2021separation,rangamani2023feature,kothapalli2024neural,qin2021contrastive}. Feature learning is a striking departure from kernel machines or random feature models (RFM) which compute linear functions of \emph{fixed} features \cite{hofmann2008kernel,rahimi2007random,daniely2016toward}. How it emerges from microscopic interactions between millions of artificial neurons is a central open question in deep learning~\cite{radhakrishnan2024mechanism,lou2022feature,baratin2021implicit,yang2021tensor,chizat2023steering}.

Even with a single hidden layer~\cite{mei2018mean,chizat2018global,wang2021hidden,arnaboldi2023high}, the interplay between initialization~\cite{luo2021phase}, width~\cite{maillard2023injectivity,pacelli2023statistical,baglioni2024predictive}, learning rate~\cite{cui2024asymptotics,sohl2024boundary}, batch size~\cite{marino2024phase,dandi2024benefits}, and data~\cite{ciceri2024inversion,boursier2024early,goldt2020modeling} results in a bewildering range of training dynamics. Deeper networks can be analyzed in various asymptotic regimes~\cite{jacot2018neural, guth2023rainbow}, with simplified  training~\cite{chan2022redunet, fang2021exploring}, or without non-linearity~\cite{arora2018convergence,li2021statistical}. There have been exciting advances in the infinite-width limit~\cite{bordelon2022self,bordelon2024dynamical} and on low-dimensional SGD dynamics~\cite{ben2022high,arous2023high}. These give invaluable insight, but the full deep, non-linear setting has eluded a statistical mechanics approach where features emerge from microscopic interactions. A change of perspective may help close this gap.

In this paper we take a thermodynamical, top-down approach and look for a simple phenomenological model which captures the feature learning phenomenology. We show that DNNs can be mapped to a phase diagram defined by noise and nonlinearity, with phases where layers learn features uniformly, and where deep or shallow layers learn better. ``Better'' is quantified through \emph{data separation}---the ratio of feature variance within and across classes. To explain this phase diagram, we propose a macroscopic theory of feature learning in deep, nonlinear neural networks: we show that the stochastic dynamics of a nonlinear spring--block chain with asymmetric friction fully reproduce the phenomenology of data separation over training epochs and layers. The phase diagram is universal with respect to the source of stochasticity: varying dropout rate, batch size, label noise, or learning rate all result in the same phenomenology.

Our findings generalize recent work showing that in many DNNs each layer improves data separation by the same factor, with surprising regularity~\cite{he2023law}. This \emph{law of data separation} can be proved for linear DNNs with a particular choice of data and initialization~\cite{yaras2023law,wang2023understanding}. It is however puzzling why just as many networks do not abide by it. FIG.~\ref{fig: deformed-maintext} shows that training the same DNN with three different parameter sets results in strikingly different distributions of data separation over layers. Even linear DNNs induce a complex energy landscape \cite{gardner1988optimal, baldassi2021unveiling, becker2020geometry,shi2024homophily,  pennington2017geometry, fernandez2022continuous, chizat2018global} and nonlinear training dynamics~\cite{li2021statistical} that can result in non-even separation. Understanding why this happens and how it affects generalization is key to understanding feature learning.

\begin{figure*}[ht!]
    \centering
    \includegraphics[width=\linewidth]{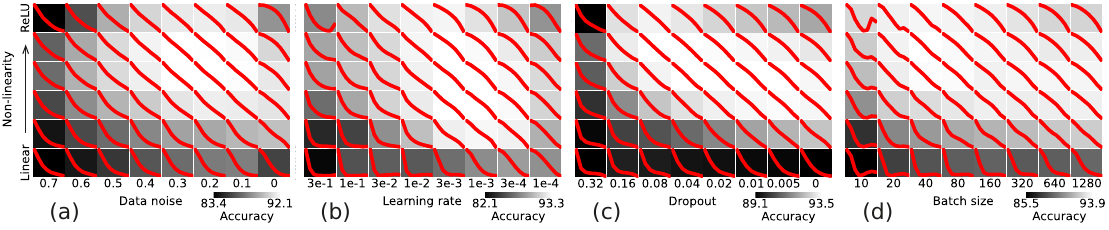}
    \caption{Phase diagrams of DNN \new{training} load curves (red) for nonlinearity vs. (a) data noise,  (b) learning rate, (c) dropout, and (d) batch size. \new{The non-linearity is controlled by the negative slope in LeakyReLU, with values of 1, 0.8, 0.6, 0.4, 0.2, and 0 from the bottom row to the top row.} In all cases, noise is strongest on the left, and nonlinearity strongest at the top. Background shading encodes test accuracy. Results are averaged over 10 independent runs on MNIST.}
    \label{fig: pd}
\end{figure*}

In our theory, spring elongations model data separation. The empirical risk exerts a load on the network to which its layers respond by separating the data, subject to nonlinearity modeled by friction. Difference in length between consecutive springs results in a load on the incident block. Friction models dynamical nonlinearity which absorbs load (or spoils the signal in gradients), causing the shallow layers to ``extend'' (separate data) less. Stochasticity from stochastic gradient descent (SGD)~\cite{yang2023stochastic,sclocchi2024different}, dropout~\cite{srivastava2014dropout}, or noisy data~\cite{sukhbaatar2014learning}, reequilibrates the load.  

The resulting model reproduces the dynamics and the phase diagram of feature learning surprisingly well. It explains when data separation is uniformly distributed across layers and when deep or shallow layers learn faster. It shows why depth may hurt and why nonlinearity is a double-edged sword, resulting in expressive models but facilitating overfitting. A stability argument suggests a link between generalization and layerwise data separation which we remarkably find enacted in real DNNs. \new{This observation is of great practical interest: together with our understanding of noise and nonlinearity it suggests a simple strategy for hyperparameter tuning and model selection which may be a compelling alternative to grid-search approaches; we explore this in FIG.~\ref{fig: CNN} and in further experiments in  SM~\cite{suppl}.}

\paragraph{Feature learning across layers of DNNs---}
A DNN with $L$ hidden layers, weights $\mW_{\ell} \in  \mathbb{R}^{d_{\ell}\times d_{\ell-1} }$, biases $\vb_{\ell} \in \mathbb{R}^{d_\ell}$, and activation $\sigma$ maps the input $\vx_0 \equiv \vx$ to the output (a label) $\vy \equiv \vx_{L+1}$ via a sequence of intermediate representations $\vx_{0} \to \vx_{1} \to \cdots \to \vx_L \to \vx_{L+1} \equiv \vy $, where $\vx_{L+1} = F(\vx) = \mW_{L+1} \vx_{L} + \vb_{L+1}$ and
\begin{equation}\label{eqn: DNN}
\begin{aligned}
    \vh_{\ell} = \mW_{\ell} \vx_{\ell-1} + \vb_{\ell}, \quad  
    \vx_{\ell} = c_{\ell}\sigma(\vh_{\ell}),
\end{aligned}
\end{equation}
for $\ell = 1, \ldots, L$. \new{We call the layers with small $\ell$ (near the input) \textit{shallow} and those with large $L$ (near the output) \textit{deep}.} The activation-dependent normalization factors $c_{\ell}$ scale the variance of hidden features in each layer close to $1$~\cite{he2015delving,poole2016exponential,schoenholz2016deep}; they can be replaced by batch normalization~\cite{ioffe2015batch}.

It is natural to expect that in a well-trained DNN the intermediate features $\vx_\ell$ improve progressively over $\ell$. Following recent work on neural collapse we measure separation as the ratio of variance within and across classes~\cite{papyan2020prevalence, he2023law, zarka2021separation,rangamani2023feature,yaras2023law}; in supplemental material (SM) \cite{suppl} we show that analogous phenomenology exists in regression.
Let $\mathcal{X}_{\ell}^{k}$ collect the $\ell$th postactivation for examples from class $k$, with $\bar{\vx}_\ell^k$ and $N_k$ its mean and cardinality. The \new{within-class and between-class} covariances are
\new{
\begin{equation}
\Sigma_\ell^\mathrm{w} := \operatorname*{Ave}_{n_k} \, (\mathrm{Cov} \mathcal{X}_{\ell}^k),
\quad
\Sigma_\ell^\mathrm{b} := \operatorname*{Cov}_{n_k}\left( \bar{\vx}_\ell^k\right)_{k=1}^K,
\end{equation}
where $\operatorname*{Ave}_{n_k}$ and $\operatorname*{Cov}_{n_k}$ are the weighted average and covariance with weight $n_k=N_K/N$}. $\Sigma^{\mathrm{b}}_{\ell}$ is the between-class ``signal'' for classification, and $\Sigma^\mathrm{w}_{\ell}$ is the within-class variability. Data separation at layer $\ell$ is then defined as
\begin{equation}\label{eqn: Dell}
D_{\ell}:= \log\left(\mathrm{Tr}\left(\Sigma_{\ell}^\mathrm{w}\right)/\mathrm{Tr} \left( {\Sigma_{\ell}^\mathrm{b}}\right)\right).
\end{equation}
The difference 
$
    d_{\ell} = D_{\ell-1}-D_{\ell}
$
represents the contribution of the $\ell$th layer.  We call the ``discrete curve'' $D_\ell$ vs. $\ell$ the \textit{load curve} in anticipation of the mechanical analogy. If indeed each layer improves the data representation the load curve should monotonically decrease.

Modern overparameterized DNNs may perfectly fit the training data for different choices of hyperparameters, while yielding different load curves. One extreme is an RFM where only the last layer learns while the rest are frozen at initialization \cite{rahimi2007random,hu2022universality}. As the entire load is concentrated on one layer, one might intuitively expect that a more even distribution results in better performance.

\paragraph{A law of data separation?---}

He and Su~\cite{he2023law} show that in many well-trained DNNs the load is distributed uniformly over layers, 
\begin{equation*}
    d_{\ell} \approx d_{\ell'} \quad \text{for} \quad 1 \leq \ell,\ell' \leq L,
\end{equation*}
giving a linear load curve~\cite{footnote1}. This can be proved in linear DNNs with orthogonal initialization and gradient flow training~\cite{arora2018optimization,yaras2023law,wang2023understanding}, but as we show below it is brittle: nonlinearity breaks the balance. Equiseparation thus requires additional ingredients; He and Su highlight the importance of an appropriate learning rate.

\paragraph{The noise--nonlinearity phase diagram---}

We show that DNNs define a family of phase diagrams such that (\romannumeral1) increasing nonlinearity results in increasingly ``concave'' load curves, with deeper layers learning better features (taking a higher load), $d_{\ell} <  d_{\ell'}$ for $\ell<\ell'$; (\romannumeral2) noise in the dynamics rebalances the load; and (\romannumeral3) increasing noise results in convex load curves, with shallower layers learning better features.

We report these findings in FIG.~\ref{fig: pd}. In all panels the abscissas measure stochasticity and the ordinates nonlinearity. We control nonlinearity by varying the slope $\alpha\in[0,1]$ of the negative part of a LeakyReLU activation, $\mathrm{LReLU}(x) := \max(\alpha x, x)$; $\alpha = 1$ gives a linear DNN, $\alpha = 0$ the ReLU. Consider for example FIG.~\ref{fig: pd}(a) where noise is introduced in labels and data (we randomly reset a fraction $p$ of the labels $\mathbf{y}$, and add Gaussian noise with variance $p^2$ to $\vx_{0}$). Without noise (upper right corner), the load curve is concave; this resembles an RFM or a kernel machine. Increasing noise (right to left) yields a linear and then a convex load curve.

The same phenomenology results from varying learning rate, dropout, and batch size (FIG.~\ref{fig: pd}(b, c, d)). Lower learning rate, smaller dropout, and larger batch size all indicate less stochasticity in dynamics. Martin and Mahoney~\cite{martin2017rethinking} call them temperature-like parameters; see also Zhou et al.~\cite{zhou2024temperature}. \new{In SM we give simple rules of thumb for trading off these different sources of noise~\cite{suppl}.}
In all cases high nonlinearity and low noise result in a concave load curve where parameters of deep layers move faster than those of shallow. Low nonlinearity and high noise (bottom-left) result in convex load curves. We observe the same behavior for other datasets (e.g., CIFAR10) and network architectures (e.g., CNNs); cf. SM~\cite{suppl}. The goal is to understand how varying these parameters transitions from non-feature learning models like RFM or NTK to models where all layers learn.

\paragraph{A spring--block theory of feature learning---}
We now show that the complete phenomenology of feature learning, as seen through layerwise data separation, is mirrored by the stochastic dynamics of a simple spring--block chain. As in FIG.~\ref{fig: spring}, we interpret $d_\ell$, the signed elongation of the $\ell$th Hookean spring, as data separation by the $\ell$th layer. Block movement is impeded by friction which models the effect of dynamical nonlinearity on gradients; noise in the force models stochasticity from mini-batching, dropout, or elsewhere.
\begin{figure}[t!]
    \centering
    \includegraphics[width=0.95\linewidth]{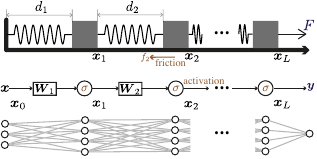}
    \caption{An illustration of the analogy between a spring--block chain and a deep neural network.}
    \label{fig: spring}
\end{figure}

The equation of motion for the position of the $\ell$th block, $x_{\ell}=\sum_{i=1}^\ell d_{i}$, ignoring block widths, is
\begin{equation}
\label{eqn: second order dynamics}
\begin{aligned}
    \ddot{x}_\ell 
    &= 
    k(x_{\ell+1} - 2x_{\ell} + x_{\ell-1}) - \gamma \dot{x}_{\ell} + f_\ell + \varepsilon  \xi_\ell,
\end{aligned}
\end{equation}
for $\ell = 1, \ldots, L$, where we assumed unit masses, $k$ is the elastic modulus, $\gamma$ is a linear damping coefficient sufficiently large to avoid oscillations, \new{$\epsilon$ controls the noise strength} and $\xi_\ell$ is noise such that $\langle \xi_\ell(t)\new{,}\xi_\ell(s) \rangle = \delta(t - s)$. As DNNs at initialization do not separate data we let $x_{\ell}(0) = 0$. The dynamics is driven by force applied to the last block $F=k(y-x_{L})$. We set $x_0 \equiv 0$ and $x_{L+1} \equiv y$ to model training data and targets. The load curve plots the distance of the $\ell$th block from the target $D_{\ell}=y-x_{\ell}$. It reflects how much the $\ell$th spring---or the $\ell$th layer---contributes to ``explaining'' the total extension---or the total data separation. 
One key insight is that the friction must be asymmetric to model the propagation of noise during training as we elaborate below. We set the sliding and maximum static friction to $\mu_\rightarrow$ for rightward and $\mu_\leftarrow$ for leftward movement. In this model the friction acts on noisy force. If noise is added to the velocity independently of the friction we obtain a more standard (but physically less realistic) Langevin dynamics which exhibits similar qualitative behavior ; for additional details see SM~\cite{suppl}.

\paragraph{The spring--block dynamics of data separation in DNNs---}

We now show experimentally and analytically that the proposed model results in a phenomenology analogous to that of data separation in real networks. There is a striking similarity between the phase diagram of the spring--block model in FIG.~\ref{fig: spring phase diagram first} and the DNN phase diagrams in FIG.~\ref{fig: pd}. Not only are the equilibria of the two systems  similar, but also the stochastic dynamics; we show this in FIG.~\ref{fig: deformed-maintext}.
\begin{figure}[t!]
    \centering
    \includegraphics[width=0.8\linewidth]{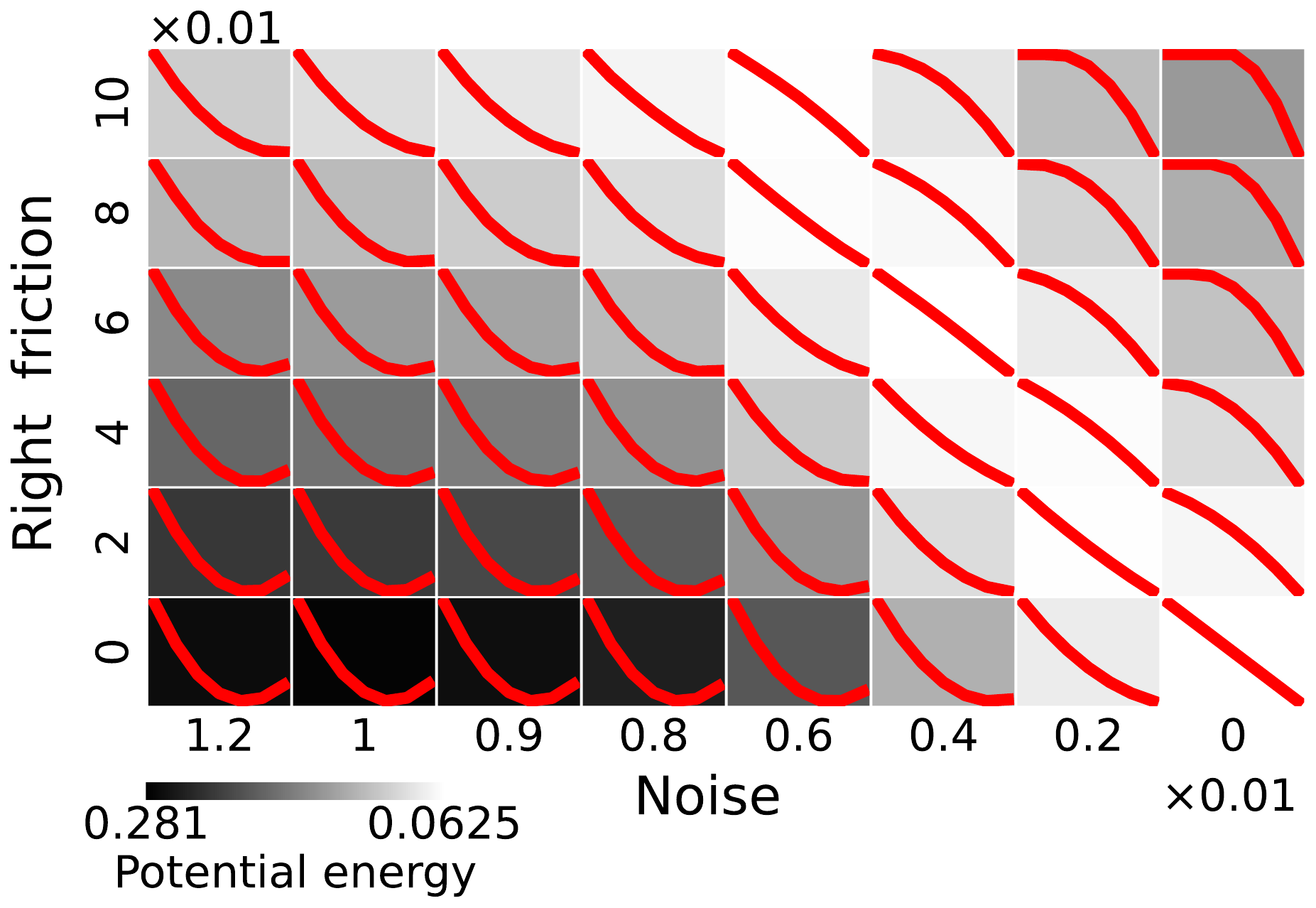}
    \caption{Phase diagram of the spring--block system \eqref{eqn: first-ord continuous} for friction $\mu_\rightarrow$ vs. noise level $\epsilon$. We set $k=1, \mu_\leftarrow=0.2$ and $L=7$. The load curves ($D_\ell=y-x_\ell$) are recorded at $t=100$; the shading corresponds to the elastic potential energy.}
    \label{fig: spring phase diagram first}
\end{figure}

\paragraph{1: Nonlinearity breaks the separation balance---}

We first show how our model explains concave load curves. For simplicity we work in the overdamped approximation $\gamma \gg 1$; in SM \cite{suppl} we show that the second order system has the same qualitative behavior. Scaling time by $\gamma$, Eq.~\eqref{eqn: second order dynamics}  yields
\begin{equation}\label{eqn: first-ord continuous}
    \dot{x}_\ell= \sigma\big(k \, (\boldsymbol{L}x)_\ell +\varepsilon \xi_\ell\big)
\end{equation}
where $(\boldsymbol{L} x)_\ell:=x_{\ell+1}-2x _{\ell}+x_{\ell-1}$ and \new{$\sigma(z) = 0$ if $-\mu_{\leftarrow} \leq z \leq \mu_{\rightarrow}$; $z - \mu_{\rightarrow}$ if $z > \mu_{\rightarrow}$; and $z + \mu_{\leftarrow}$ if $z < -\mu_{\leftarrow}$}.

Without noise and friction ($\varepsilon = 0$,  $\mu_\leftarrow = \mu_\rightarrow=0$) the system is linear with the trivial unique equilibrium 
$ 
    d_{\ell}^* \equiv y/(L+1)
$
for all $\ell$, which corresponds to the state of lowest elastic potential energy. However, analyzing the resulting system of ODEs shows that adding friction in \eqref{eqn: first-ord continuous} immediately breaks the symmetry and results in an \emph{unbalanced} equilibrium,
\begin{equation}\label{eqn: no noise ode star}
    x_{\ell}^* =\frac{y \ell}{L+1}- \frac{\mu_{\rightarrow}}{k}\left(\frac{L \ell}{2}-\frac{\ell \left(\ell-1\right)}{2}\right)
\end{equation}
if we assume the initial elastic potential $k(y-x_0)^2/2$ is sufficiently large such that all blocks eventually move. In this case $\Delta d^*_\ell := d^*_{\ell+1} - d^*_{\ell} =(\mL x^*)_\ell = {\mu_\rightarrow}/{k} >0$ and the load curve is concave. Note that this result only involves $\mu_\rightarrow$ but not $\mu_\leftarrow$ as without noise and with sufficient damping the blocks only move to the right. The interpretation is that in equilibrium, the friction at each block absorbs some of the load so that the shallower springs extend less; in a DNN this corresponds to a regime where the deepest layer, whose gradients do not experience nonlinearity, does most of the separation.

\begin{figure}[t!]
    \centering
    \includegraphics[width=0.95\linewidth]{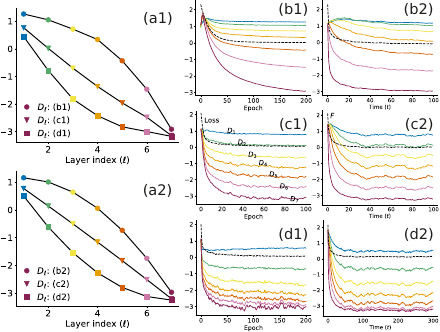}
    \caption{The load curves at convergence (\textbf{a}) and trajectories (\textbf{b}, \textbf{c}, \textbf{d}) for a $7-$hidden layer ReLU MLP on MNIST (\textbf{\rule[0.0ex]{0.2cm}{0.4pt}1}) vs our spring--block model (\textbf{\rule[0.0ex]{0.2cm}{0.4pt}2}). For the MLP (\textbf{\rule[0.0ex]{0.2cm}{0.4pt}1}), the ordinate is \(D_\ell\) (data separation at layer $\ell$); the dashed line is the training loss. Characteristic behaviors: (b) high nonlinearity (high friction) and low randomness in training (noise in force); (c): balanced nonlinearity and randomness; (d) low nonlinearity and strong randomness.
    In the spring system  (\textbf{\rule[0.0ex]{0.2cm}{0.4pt}2}), the ordinate is the distance to the target $D_\ell=y-x_\ell$. The values are scaled to match the same regime used in the DNN for ease of visualization. The dashed line is the force at the rightmost block $F$ which is at a different scale from $D_\ell$.} 
    \label{fig: deformed-maintext}
\end{figure}

\paragraph{2: Noise reequilibrates the load---}
If friction reduces load, how can  a chain with friction---or a nonlinear DNN---result in a uniformly distributed load? We know from FIG.~\ref{fig: pd} that in DNNs stochastic training helps achieve this. We now show how our model reproduces this behavior and in particular a counterintuitive phenomenon in FIG.~\ref{fig: pd} where large noise results in \emph{convex} load curves where shallow layers learn better than deep. We show that this happens when $\mu_{\leftarrow} > \mu_{\rightarrow}$.

We begin by defining the \emph{effective} friction over a time window of length $\eta$. Let $\overline{\varepsilon} := \varepsilon / \sqrt{\eta}$ and $\zeta_t \sim (\xi_{t + \eta} - \xi_t)/\sqrt{\eta}$ (iid). We will assume for convenience that the increments $\zeta$ are bounded (see SM~\cite{suppl} for details). The effective friction is then
\begin{equation*}
    \mu_{\text{eff}}
    :=
    \mathbbm{E}_{\zeta}\big[\left(\sigma(k\Delta d+\overline{\varepsilon}\zeta)-k\Delta d\right) \mid \sigma(k\Delta d+\overline{\varepsilon} \zeta) \neq 0\big].
\end{equation*}
For large noise we will have that $ \mu_{\text{eff}} \approx\lim_{\overline{\varepsilon} \to \infty} \mu_{\text{eff}} = \frac{1}{2}(\mu_\rightarrow - \mu_\leftarrow)$. Since this effective friction is approximately independent of $x$, the load curve can be approximated by substituting $\mu_{\text{eff}}$ for $\mu_\rightarrow$ in Eq.~\eqref{eqn: no noise ode star}, which leads to 
\begin{equation}
\label{eqn: d infty star}
\Delta d^*  \approx \frac{1}{k} \lim_{\overline{\varepsilon} \to \infty} \mu_{\text{eff}}(\overline{\varepsilon}) = \frac{\mu_\rightarrow - \mu_\leftarrow}{2k}.
\end{equation}
It is now clear that with sufficient noise the load curve is concave if $\mu_{\rightarrow}>\mu_{\leftarrow}$, linear if $\mu_{\rightarrow}=\mu_{\leftarrow}$, and convex if $\mu_{\rightarrow}<\mu_{\leftarrow}$. We can also see that for $\overline{\varepsilon} = 0$, $\mu_\text{eff} = \mu_\rightarrow$, so that this effective friction correctly generalizes the noiseless case Eq.~\eqref{eqn: no noise ode star} and we have that $\Delta d^* \in [\frac{\mu_\rightarrow-\mu_\leftarrow}{2k}, \frac{\mu_\rightarrow}{k}]$.

Further, when $\Delta d \in [\frac{\mu_\rightarrow-\mu_\leftarrow}{2k}, \frac{\mu_\rightarrow}{k}]$, it holds that $\frac{\mathrm{d} \mu_{\text{eff}}} {\mathrm{d} \varepsilon}\leq 0$. It implies that increasing noise always decreases effective friction. This resembles phenomena like acoustic lubrication or noise-induced superlubricity \cite{shi2021micro}. When $\mu_\leftarrow>\mu_\rightarrow>0$, as we vary $\varepsilon$ from $0$ to $\infty$ we will first see a concave, then a linear, and finally a convex load curve. Therefore, when $\mu_\leftarrow>\mu_\rightarrow>0$ our model explains the entire DNN phase diagram.

How can we relate the condition $\mu_\leftarrow>\mu_\rightarrow>0$ to DNN phenomenology? Note that in our model $\mu_\leftarrow$ is activated only due to the noise, since the signal (the pulling force) is always to the right. In a DNN, the forward pass of the backpropagation algorithm computes the activations while the backward pass computes the gradients. In the forward pass, the input is multiplied by the weight matrices starting from the shallowest to the deepest. With noise (e.g., dropout), the activations of the deepest layer accumulate the largest noise. Thus noise in DNN training chiefly propagates from shallow to deep, yielding noisiest gradients in deepest layers. This is exactly what $\mu_\leftarrow > \mu_\rightarrow$ in our model implies, since $\mu_\leftarrow$ is only triggered by noise. 

Indeed, from Eq.~\eqref{eqn: d infty star} we see that when only friction for leftward movement is present, $\mu_\leftarrow > \mu_\rightarrow = 0$, the load curve is convex if $\varepsilon > 0$ and linear if $\varepsilon=0$ (FIG.~\ref{fig: orth}(b)). The equilibrium is independent of $\mu_\leftarrow$ since no block moves left. This parallels findings in linear DNNs where training with gradient flow and ``whitened'' data leads to a linear load curve~\cite{yaras2023law}. In contrast, introducing noise makes the load curve convex (FIG.~\ref{fig: orth}(a)). This shows that \emph{dynamical} nonlinearity---or friction in our model---exists even in linear DNNs, so that their learning dynamics are nonlinear~\cite{li2021statistical}.
\begin{figure}[t!]
    \centering
    \includegraphics[width=0.9\linewidth]{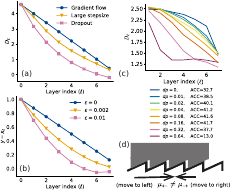}
    \caption{(a) Load curve of a deep linear neural network on random orthonormal datasets. (b) Spring--block model without left friction ($\mu_\rightarrow = 0$) with $\mu_\leftarrow=0.1$, under varying levels of noise $\varepsilon$. \new{(c) Load curve of 7-layer CNN trained on the CIFAR10 dataset; \textit{dp} and ACC denote the dropout ratio and test accuracy.} (d) An illustration of asymmetric friction.}
    \label{fig: orth}
\end{figure}

\paragraph{3: Equiseparation minimizes elastic potential energy and improves generalization---} 

We finally show how our theory gives insight into generalization. Among all spring--block chains under a fixed load, the equiseparated one is the most stable in the sense of having the lowest potential energy. This motivates us to study the test accuracy of DNNs---shown as background shading in FIG.~\ref{fig: pd}---as a function of load curve curvature. The result is intriguing: linear load curves correspond to the highest test accuracy. It suggests that by balancing nonlinearity with noise, DNNs are at once highly expressive and not overfitting. That a uniform load distribution yields the best performance is intuitively pleasing, but it is also valuable operationally as it may help guide training to find better networks. \new{In FIG.~\ref{fig: CNN} we show a proof-of-concept example with a CNN: initial training yields a concave load curve. Our theory suggests that more noise in training would flatten the load curve and improve generalization---which is indeed what happens. Further details and experiments with very deep networks can be found in SM~\cite{suppl}. }

\paragraph{Conclusion---}
Deep learning theories are mostly built bottom-up, but fields like physics, biology, neuroscience and economics benefit from both bottom-up and top-down, phenomenological approaches. Mechanical analogies such as spring-block models play an important role across science; a prime example is the Burridge--Knopoff model in seismology~\cite{burridge1967model} and related ideas in neuroscience~\cite{hopfield1994neurons}. We think that deep learning can similarly benefit from both paradigms, especially with complex phenomena like feature learning which require to simultaneously consider depth, nonlinearity, noise and data.

Our phenomenological model elucidates the role of nonlinearity and randomness and suggests exciting connections with generalization \new{that may inspire new theory, but also, importantly, new approaches to hyperparameter tuning and model selection. The current theory  applies to DNNs where inner layers are alike; extensions to  heterogeneous DNNs will require refined definitions of data separation and new mechanical models.} 

We finally mention that we considered various cascade structures as possible analogies to DNNs; the SM links to real experiments with a folding ruler~\cite{suppl}. 

\begin{figure}[t!]
    \centering
    \includegraphics[width=0.95\linewidth]{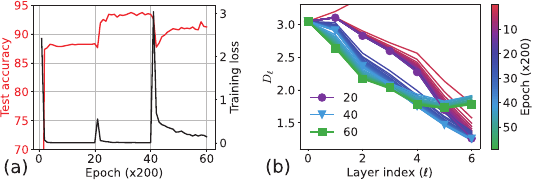}
    \caption{\new{The dynamics of load curves for a deep CNN. (a) Test accuracy versus training loss. (b) The corresponding load curves during training. In the experiments, we introduce $5\%$ at epoch $20\times 200$ and $30\%$ dropout at epoch $40\times 200$.}}
    \label{fig: CNN}
\end{figure}

\paragraph{Acknowledgments---} We are grateful to Hangfeng He,  Weijie Su and Qing Qu for inspiring discussions about data separation. Shi and Dokmanić were supported by the ERC Starting Grant 852821—SWING. Pan acknowledges support from National Natural Science Foundation of China (NSFC), Grant No.~62006122 and 42230406.

\bibliography{ref}
\clearpage

\appendix

\renewcommand{\thefigure}{S\arabic{figure}}
\setcounter{figure}{0}

\renewcommand{\theequation}{S\arabic{equation}}
\setcounter{equation}{0}

\title{Spring–block theory of feature learning in deep neural networks \\ SUPPLEMENTAL MATERIAL}

\maketitle

\section{S1. A folding ruler experiment}

\begin{figure}[ht!]
    \centering
    \includegraphics[width=\linewidth]{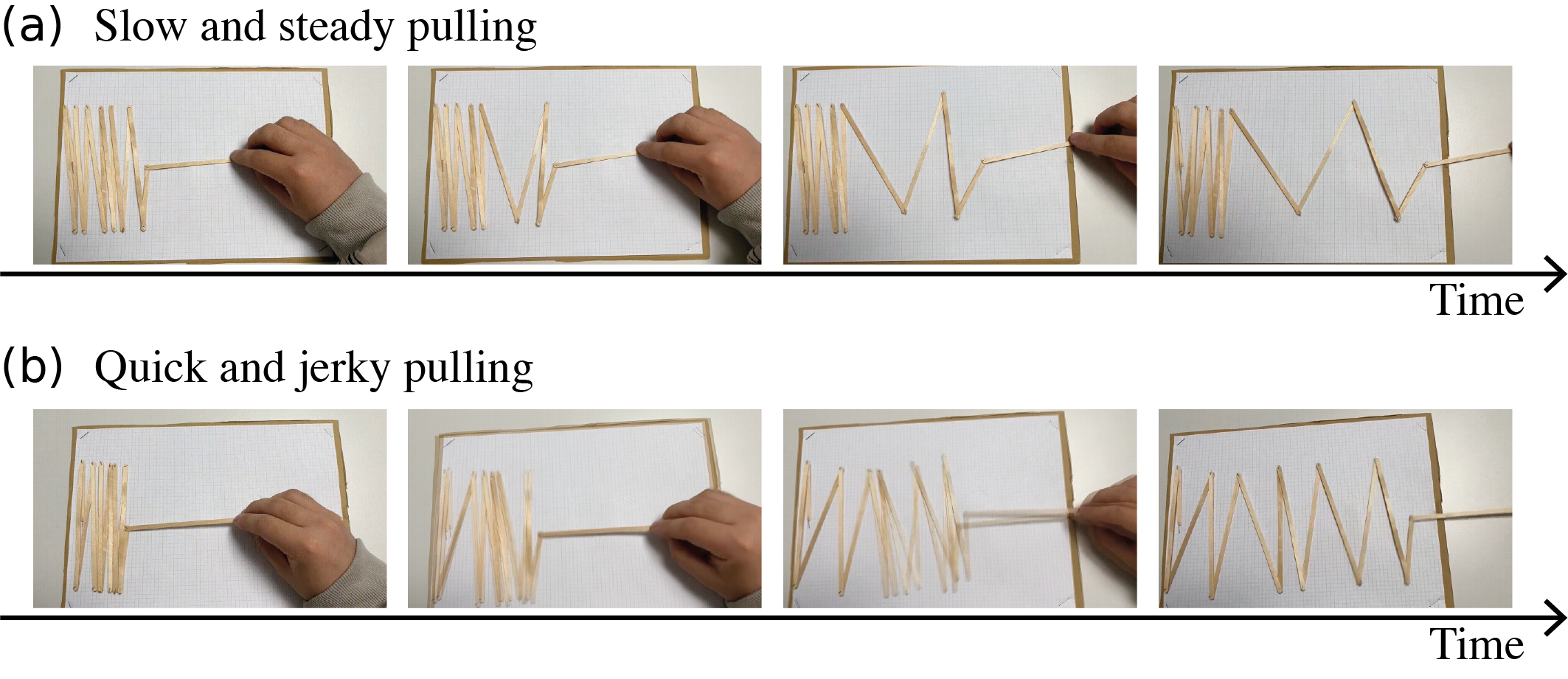}
    \caption{Two ways to extend a folding ruler: (a) With slow and steady pulling, the outer layer extends more than the inner layer due to friction. (b) With quick and jerky pulling, the ``active'' dynamics redistribute the extension across the different layers.}
    \label{fig: folding ruler}
\end{figure}

We first describe an experiment where a different cascaded mechanical system, a folding ruler, exhibits a phenomenology which is in some ways reminiscent of feature learning dynamics in DNNs. We emphasize that this reminiscence is anecdotal but nonetheless mention it since it motivated our work. The goal of the experiment is to show that noise can renegotiate the imbalance caused by friction. As shown in FIG.~\ref{fig: folding ruler}, we pin the left end of the folding ruler and pull the right end by hand. Due to friction, if we pull it very slowly and steadily, the outer layer extends far while the inner layers are close to stuck. This reminds us of lazy training where the outer layers take the largest proportion of the load. Conversely, shaking while pulling helps ``activate'' the inner layers and redistribute the force, and ultimately results in a uniformly distributed extension of each layer. Videos can be found at \url{https://github.com/DaDaCheng/DNN_Spring}

\section{S2. Additional experimental results on various networks and datasets}

\paragraph{Convolutional neural networks---} Similarly to MLPs, deep convolutional neural networks (CNNs) also learn features and progressively collapse data. Here we explore how the observation that a uniform load distribution correlates with better generalization may help guide CNN training. In FIG.~\ref{fig: CNN} of the main text, we illustrate an experiment where we first train a CNN using Adam~\cite{kingma2014adam} at a learning rate of $10^{-4}$ on MNIST. We apply both pooling and upsampling with the same ratio after each activation function to ensure that all intermediate features have the same size. Training converges approximately after $20\times 200$ epochs, resulting in a concave load curve (purple). Next, we introduce a $5\%$ dropout which causes the training to resume and the load curve to become linear (blue). Importantly, it also improves accuracy, as predicted by the theory. Stronger noise (a $30\%$ dropout at $40 \times 200$ epochs) at once results in a convex load curve (green curve) and worse generalization. 

\begin{figure}[h!]
    \centering
    \includegraphics[width=0.85\linewidth]{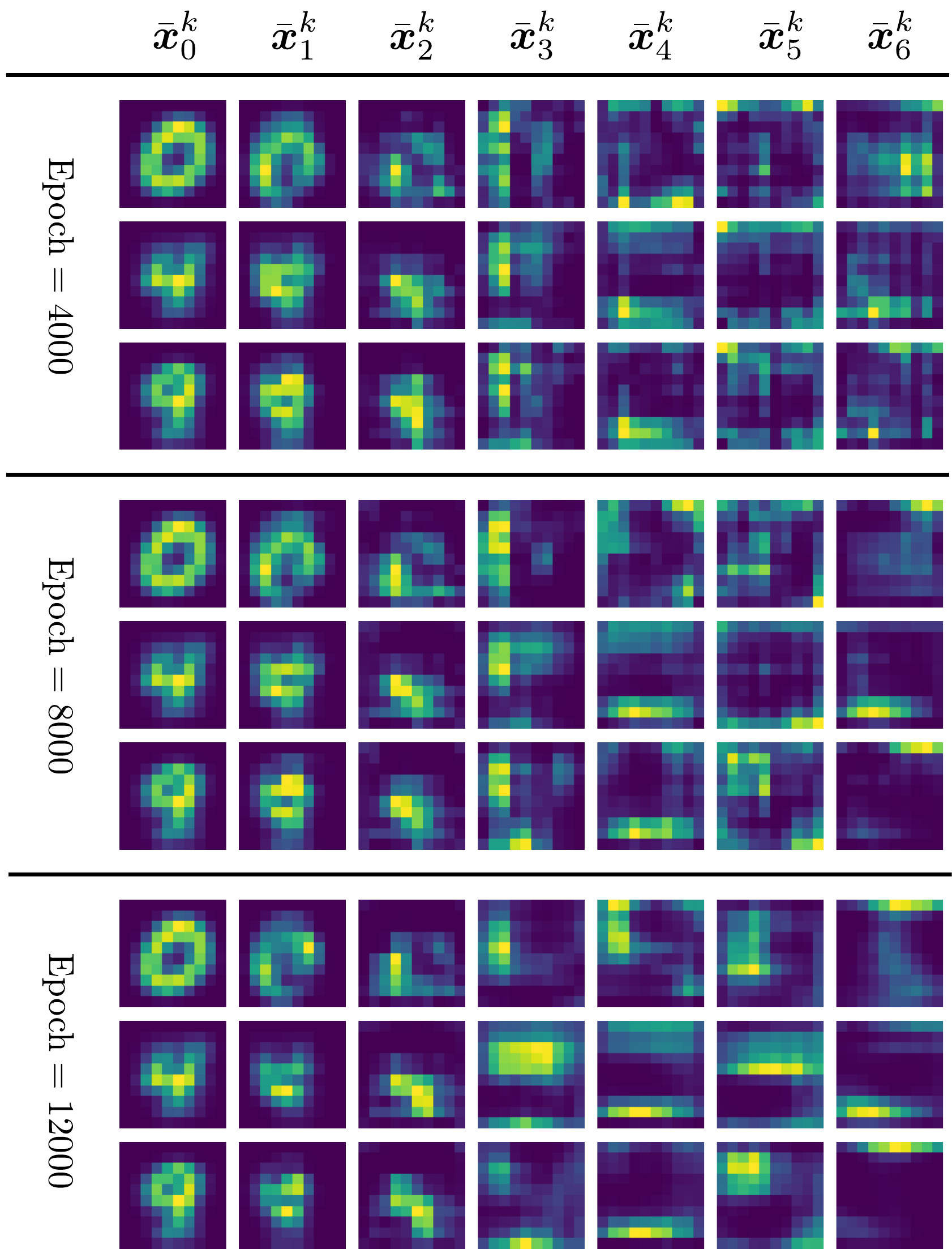}
    \caption{In-class means of the mid layer features in deep CNN whose training dynamics is showed in FIG.~\ref{fig: CNN}. We plot first channel at each layer for digits 0, 4, and 9 at epochs 4000 (concave load curve), 8000 (linear load curve), and 12000 (convex load curve).}
    \label{fig: CNN_details}
\end{figure}

In FIG.~\ref{fig: CNN_details}, we illustrate the in-class means of the $\ell$-th mid-feature $(\bar{\vx}_\ell^k)$ at epochs 4000, 8000, and 12000. The corresponding load curves at these times are concave, linear, and convex, respectively, as shown in FIG.~\ref{fig: CNN}.

\begin{figure}[t!]
    \centering
    \includegraphics[width=0.75\linewidth]{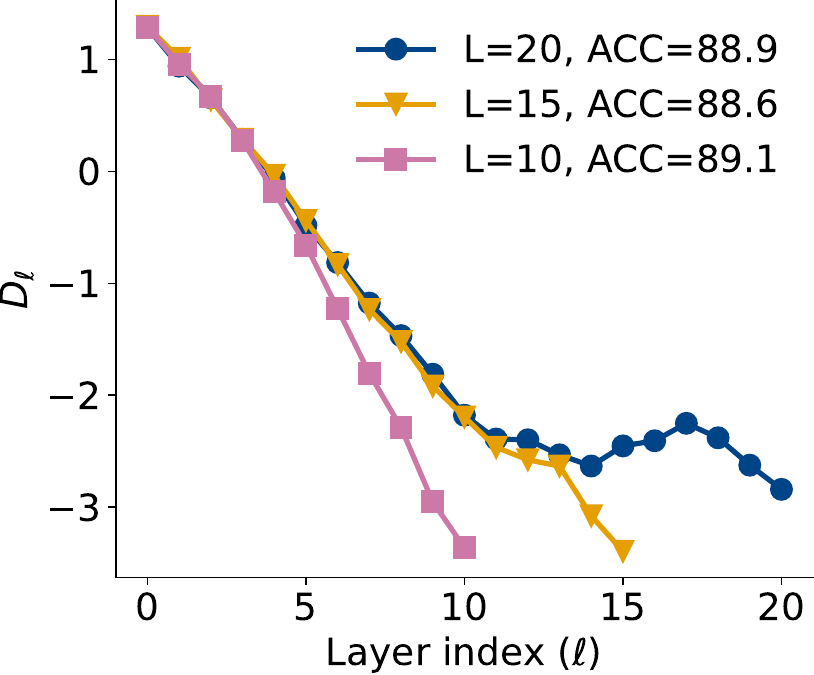}
    \caption{Load curve with different depths $L$ in ReLU MLPs.}
    \label{fig: depths}
\end{figure}

\paragraph{Depth and ``dead'' layers---} In FIG.~\ref{fig: depths}, we vary the depth of the MLP while keeping all other hyperparameters fixed. We observe that when the network is very deep, for example with $L = 20$, the separation does not go further, and even degrades, as data passes through the 14th to 18th layers. This suggests that these layers are inactive (hence ``dead'') and do not contribute to the overall task. Therefore, they can potentially be pruned without adversely affecting network performance. In this experiment for $L=20$, we pruned the ``dead'' trained layer for $\ell>13$ and added a new final linear layer, which was then retrained. This resulted in a test accuracy of 89.6\%. When we retrained all layers along with the newly added final linear layer, the test accuracy increased to 90.1\%.

\begin{figure}[h!]
    \centering
    \includegraphics[width=0.8\linewidth]{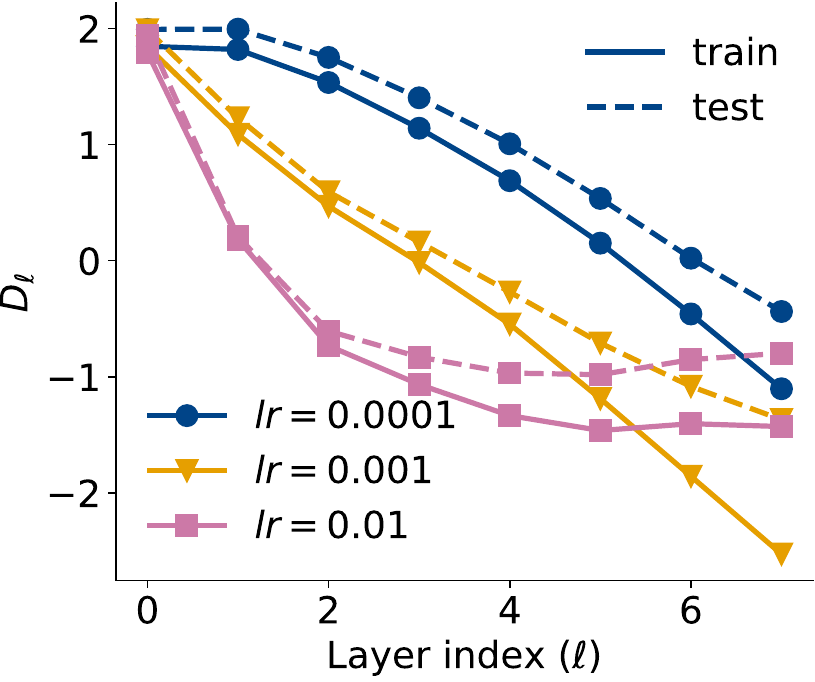}
    \caption{DNN load curves for the training set (solid line) and test set (dashed line) under different learning rates ($lr$). The corresponding test accuracies are 91.93\%, 92.52\%, and 89.99\% for learning rates of 0.0001, 0.001, and 0.01, respectively. The DNN in this experiment has a width of 784. These results were obtained from a single instance using $\alpha=0.2$ for LeakyReLU and were not averaged across multiple trials.}
    \label{fig: train_test}
\end{figure}

\begin{figure*}[t]
    \centering
    \includegraphics[width=\linewidth]{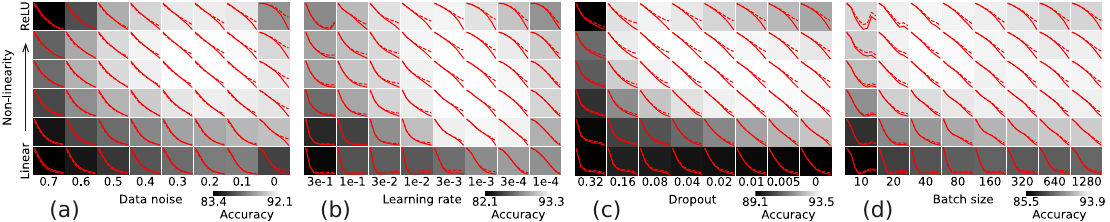}
    \caption{Phase diagrams of DNN training load curves (solid red) and test load curves (dashed red) of the experiments in FIG.~\ref{fig: pd} shown together.}
    \label{fig: pd_train_test}
\end{figure*}

\begin{figure*}[t]
    \centering
    \includegraphics[width=\linewidth]{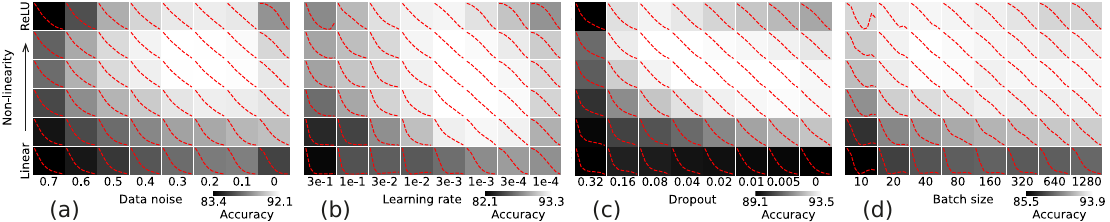}
    \caption{Phase diagram of test load curve of the experiments in FIG.~\ref{fig: pd}.}
    \label{fig: pd_test}
\end{figure*}

\paragraph{Training vs. test load curves---} This work primarily investigates the concavity and convexity of the training load curve. Notably, the test load curve exhibits behavior similar to that of the training load curve. In FIG.~\ref{fig: train_test} we show the load curves for both the training samples (solid line) and test samples (dashed line) on the MNIST dataset.  We can see that with a small learning rate, both curves are concave. Increasing the learning rate makes them both linear, and eventually convex. The most linear curve corresponds to  the highest test accuracy. The same plot as FIG.~\ref{fig: pd} but with the test load curve is shown in  FIG.~\ref{fig: pd_test}, and both train and test load curves are shown together in FIG.~\ref{fig: pd_train_test}.

\paragraph{Correlation between load linearity and generalization---}

\begin{figure}[th!]
    \centering
    \includegraphics[width=1\linewidth]{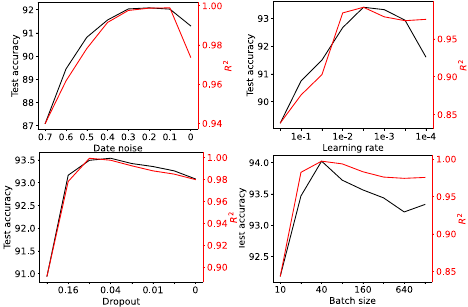}
    \caption{Comparison between test accuracy (left axis, black) and the $R^2$ coefficient of determination regression score (right axis, \textcolor{red}{red}). The data correspond to the row which contains the highest test accuracy for each phase diagram in FIG. 1 (\(\alpha=0.4\) for data noise and learning rate, and \(\alpha=0.2\) for dropout and batch size).}
    \label{fig: R2}
\end{figure}

In FIG.~\ref{fig: R2}, we report the coefficient of determination ($R^2$) to assess deviations from linearity of the load curve in FIG.~\ref{fig: pd} in the main text. We observe that the highest accuracy is always accompanied by the highest $R^2$ score. In all four phase diagrams shown in FIG.~\ref{fig: pd}, the best test accuracy occurs around the second or third row ($\alpha = 0.2$ or $\alpha = 0.4$).  Therefore, we compare the test accuracy with the $R^2$ coefficient of determination regression score in FIG.~\ref{fig: R2}.

\begin{figure}[h!]
    \centering
    \includegraphics[width=1\linewidth]{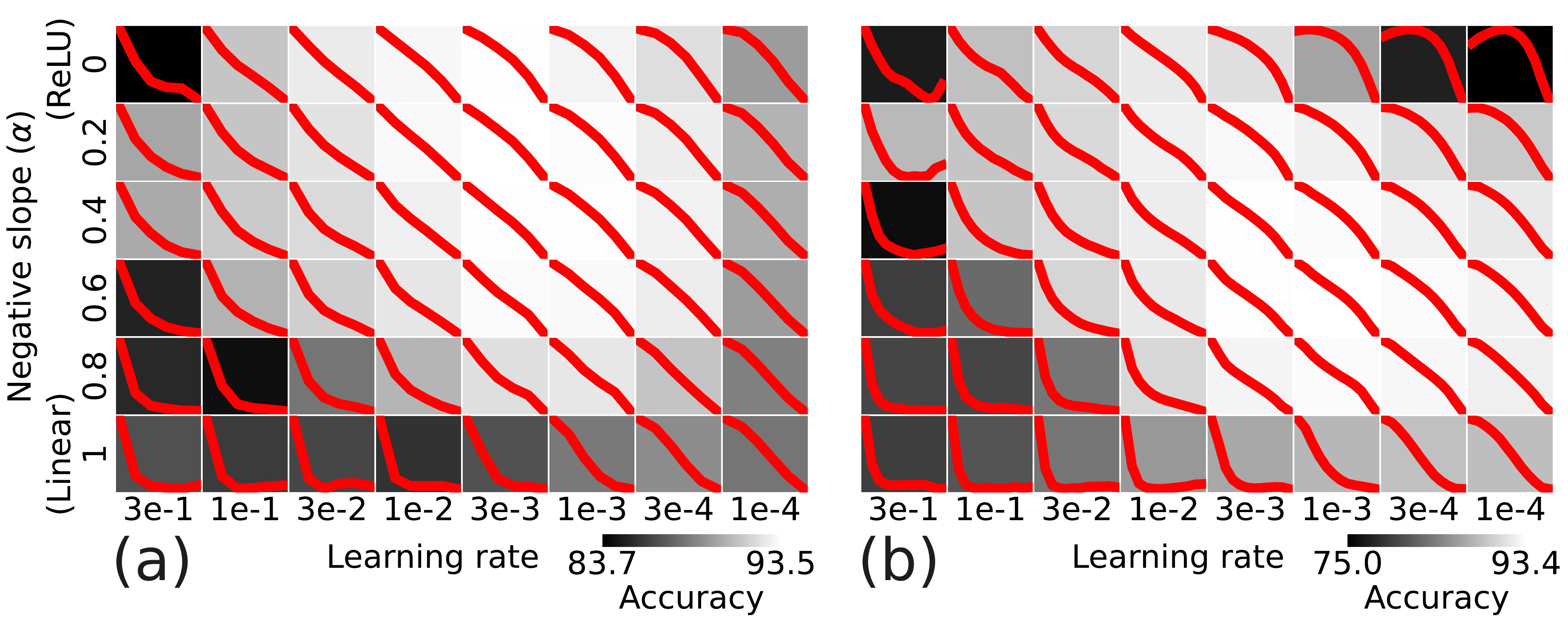}
    \caption{Phase diagram of the DNN load curves on the MNIST dataset for nonlinearity vs. learning rate at different depths: (a) 6-layers DNN, (b) 12-layers DNN.}
    \label{fig: pd_depth}
\end{figure}

\begin{figure}[h!]
    \centering
    \includegraphics[width=1\linewidth]{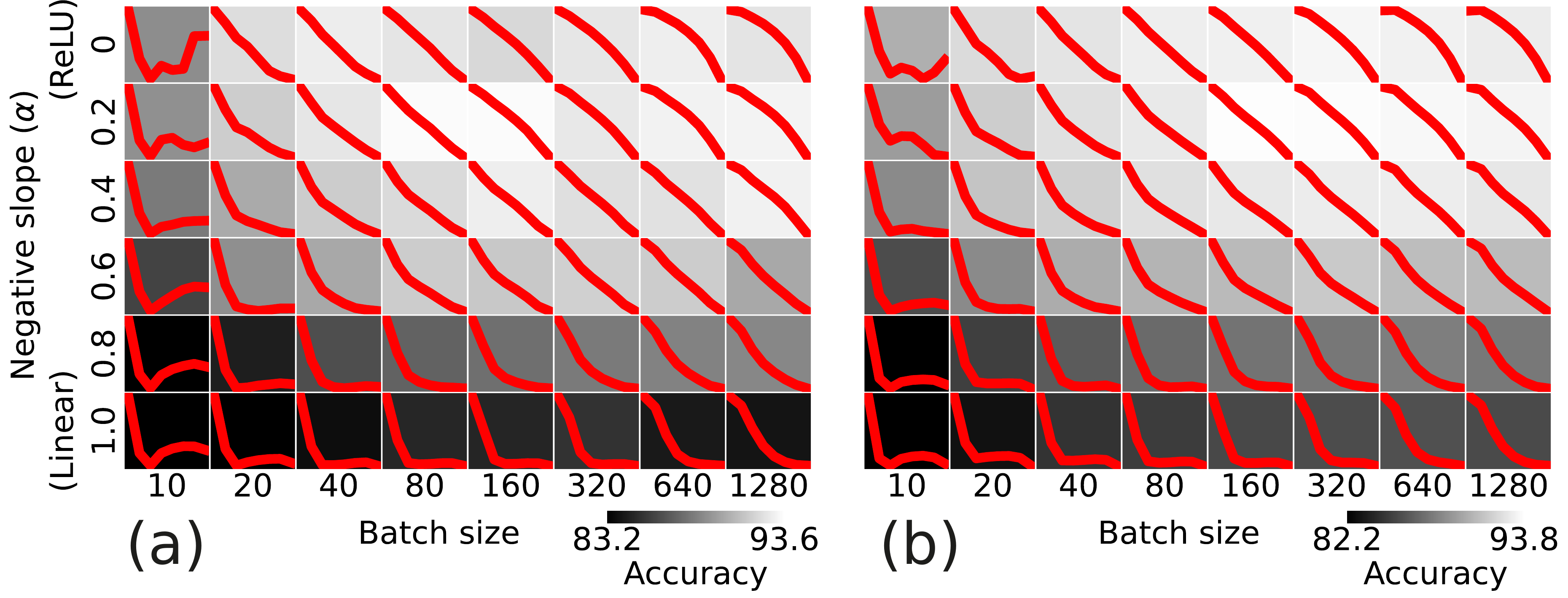}
    \caption{Phase diagram of the DNN load curves on the MNIST dataset for nonlinearity vs. batch size at different widths: (a) DNN with a width of 400, trained with the learning rate of 0.0005, (b) DNN with a width of 2500, trained with the learning rate of 0.0001.}
    \label{fig: pd_width}
\end{figure}

\begin{figure}[t!]
    \centering
    \includegraphics[width=1\linewidth]{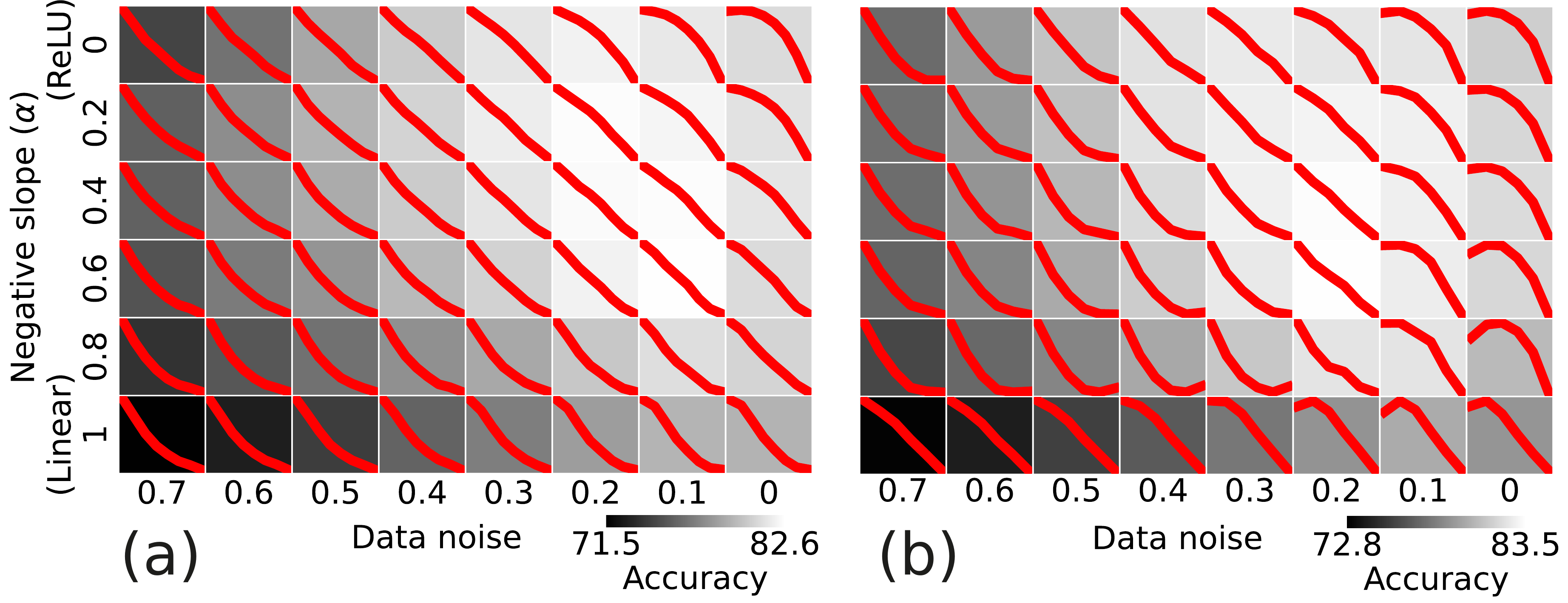}
    \caption{Phase diagram of the DNN load curves on the FashionMNIST dataset for nonlinearity vs. data noise. (a) uses the same settings as FIG.~\ref{fig: pd}. (b) shows the results for a 6-layer DNN trained with the SGD optimizer (instead of Adam) using a learning rate of 0.001. For SGD, all cases are trained for 1000 epochs. Linear DNNs (the bottom line) do not converge well with the SGD optimizer.}
    \label{fig: FashionMNIST}
\end{figure}

\begin{figure}[t!]
    \centering
    \includegraphics[width=1\linewidth]{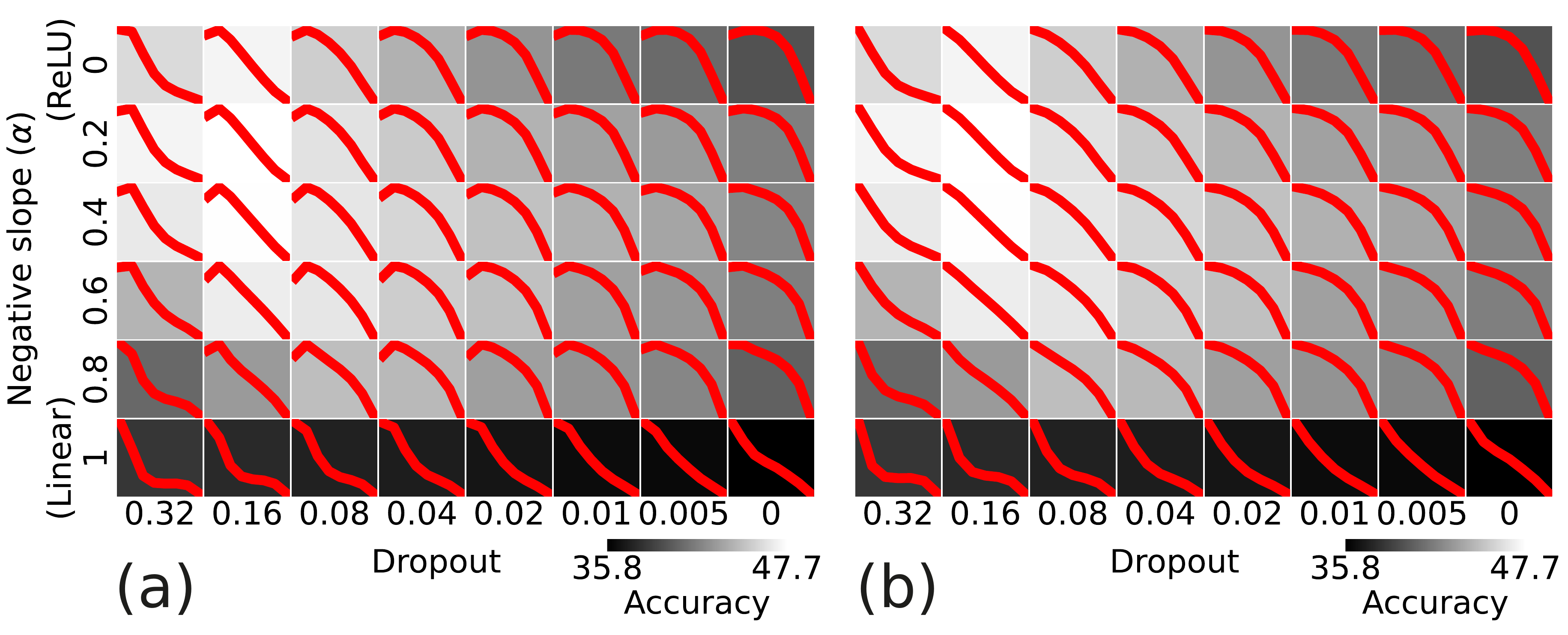}
    \caption{Phase diagram of the DNN load curves on the CIFAR10 dataset for nonlinearity vs. dropout. The red curves in (a) show $D_\ell$ vs. $\ell$ for $\ell=0,1,\cdots,7$, while (b) presents the same data as in (a) but plots $\ell$ starting from $1$ without input data $(0,D_0)$.}
    \label{fig: cifar10}
\end{figure}

\begin{figure}[t!]
    \centering
    \includegraphics[width=1\linewidth]{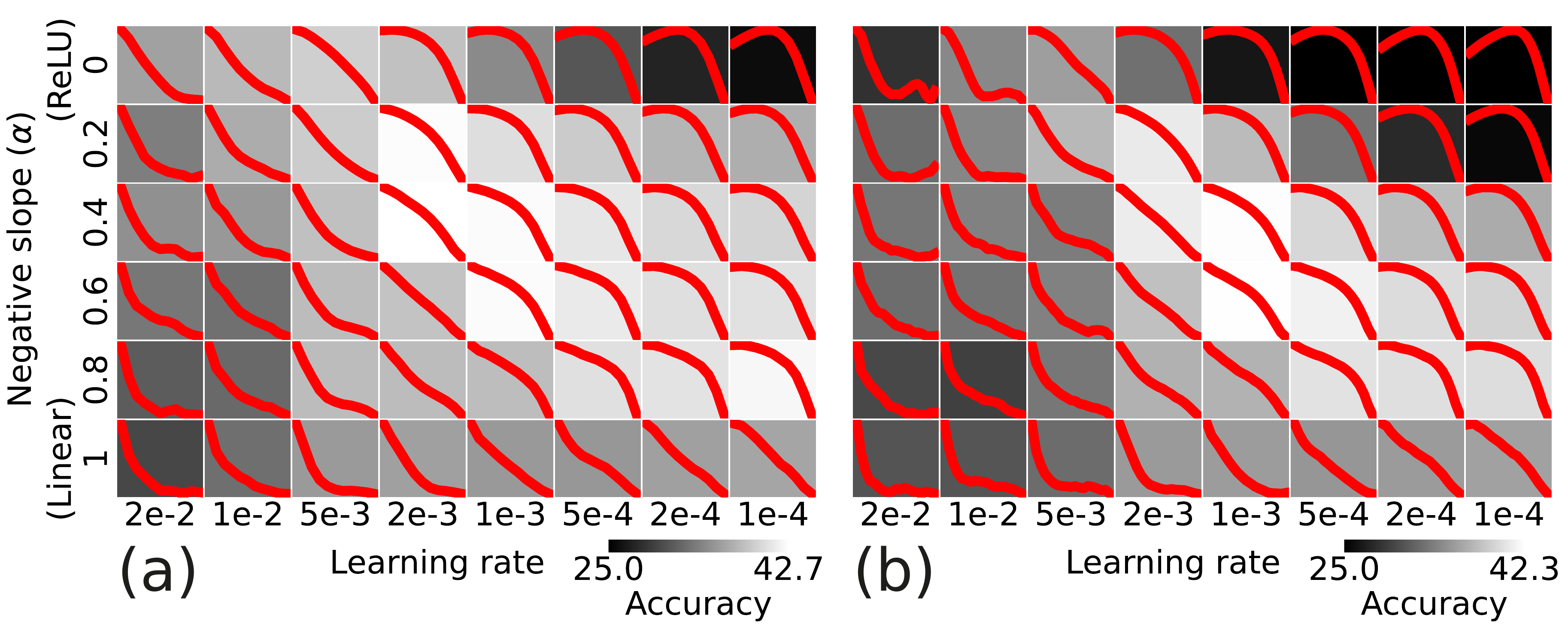}
    \caption{Phase diagram of the DNN load curves on the CIFAR10 dataset for nonlinearity vs. learning rate at different learning rate: (a) 12-layers DNN, (b) 20--layers DNN. The red curves plot $D_\ell$ staring from $1$ without input data $(0,D_0)$.}
    \label{fig: cifar10-depth}
\end{figure}

The phenomenon that the highest accuracy occurs when the load curve is linear is universally observed across a wide range of parameters and datasets. It is observed, for example, for different depths in FIG.~\ref{fig: pd_depth} and different widths in FIG.~\ref{fig: pd_width}.
It also occurs with other datasets and optimizers. We experiment with the FashionMNIST dataset in FIG.~\ref{fig: FashionMNIST} with the same default setting in (a) and a different optimizer in (b); further experiments on CIFAR10 (flatten) are shown in FIG.~\ref{fig: cifar10}. The original color images are first resampled to a size of $3\times10 \times 10$ for computational reasons and then vectorized as input to an MLP. We note that $D_0$ (the data separation metric in the input layer) slightly deviates from linearity, likely due to the additional complexity of CIFAR10, but the subsequent layers closely follow the regular phenomenology discussed in the main text (FIG.~\ref{fig: cifar10}(b)).
As shown in the experiments in FIG.~\ref{fig: cifar10}(b),  noise vs. nonlinearity phase diagram remains approximately valid even when the data becomes more complex. A small deviation from this occurs when dropout is replaced by learning rate on CIFAR10: in this case, the best generalization corresponds to slightly concave load curves (FIG.~\ref{fig: cifar10-depth}). Since learning rate is qualitatively very different from standard ``noise'', we expect that at some point it exhibits a refined phenomenology; this particular example is an intriguing avenue for further exploration.

\begin{figure}[t!]
    \centering
    \includegraphics[width=0.8\linewidth]{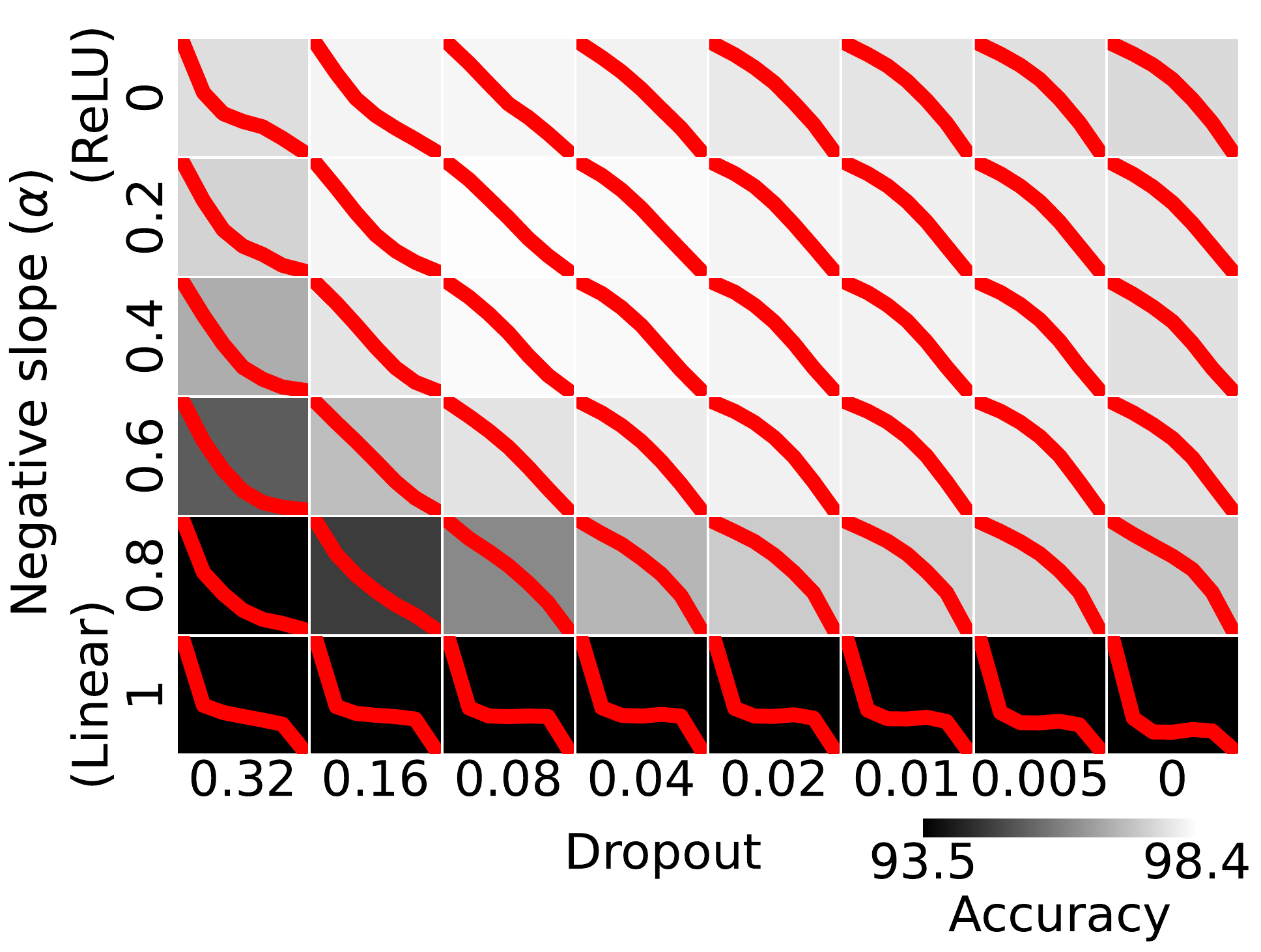}
    \caption{Phase diagram for dropout vs. non-linearity. The red curve shows the load curve for $\ell$ vs. $D_\ell$ for $\ell = 1, \dots, 7$. The background color represents the test accuracy. The results are obtained by training on the entire 60,000 training samples in the MNIST dataset.}
    \label{fig: pd_dp_full}
\end{figure}

FIG.~\ref{fig: pd} and all these phase diagrams above are trained on a small portion of the dataset mainly due to computational cost. In FIG.\ref{fig: pd_dp_full}, we trained on the same MLP on the full MNIST dataset; as these experiments show, the results are consistent with the results obtained on the downsampled dataset.

\begin{figure}
    \centering
    \includegraphics[width=\linewidth]{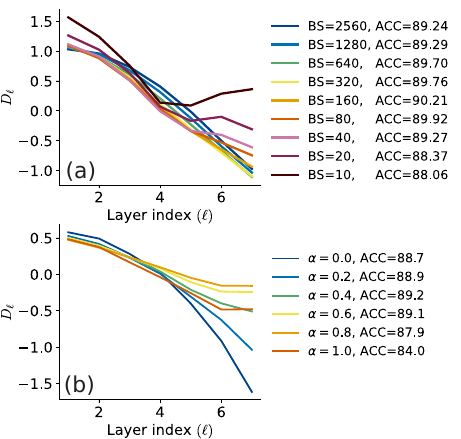}
    \caption{Load curve of a 8-layer and 400-width MLP on trained on entire FashionMNIST dataset. (\textbf{a}) Variation with different batch sizes (BS) using LeakyReLU ($\alpha=0.4$) activation, where ACC denotes test accuracy. (\textbf{b}) Variation with LeakyReLU activation with full batch for different negative slope $\alpha$. }
\label{fig: horizontal and vertical}
\end{figure}

We also train the MLP on the entire FashionMNIST dataset in FIG.~\ref{fig: horizontal and vertical}. In FIG.~\ref{fig: horizontal and vertical}(a), we fixed non-linearity and only varied batch size, while in FIG.~\ref{fig: horizontal and vertical}, we used full batch training with learning rate as $0.0001$. In both cases, the best accuracies align with the flattest load curve.

Additionally, we observe that as the training dataset size increases, it is preferable to use a wider MLP.  With larger datasets, the load curve is more likely to become wavy or twisting (neither concave nor convex). Furthermore, we find that dropout is more effective in flattening these wavy curves compared to other noise-based methods.

\section{S3. Details of numerical experiments and reproducibility}\label{sec: experiment details}
In Figs.~\ref{fig: pd}, \ref{fig: deformed-maintext}, \ref{fig: depths}, \ref{fig: train_test}, \ref{fig: pd_depth},  \ref{fig: pd_width},  \ref{fig: FashionMNIST}, \ref{fig: cifar10}, \ref{fig: cifar10-depth}, \ref{fig: pd_dp_full}, \ref{fig: horizontal and vertical}, \ref{fig: hat_D}, \ref{fig: diffD}, \ref{fig: regression} and \ref{fig: regression_MNIST}, the networks are $8$-layer fully connected MLPs (7 hidden layers), with layer width equal to $100$ unless stated otherwise. We use ReLU activations, BatchNorm in each layer, and no dropout as the default setting. All parameters are initialized using the default settings in PyTorch.  Unless otherwise specified, the networks are trained using the ADAM optimizer on 2560 training samples from the MNIST dataset, with cross-entropy loss for classification tasks. The learning rate is set to be 0.001 and the batch size is 2560 unless otherwise stated. In the phase diagrams, the load curves and test accuracies are measured on the full 10,000-sample test dataset after 100 epochs of training for the experiments on data noise, dropout, and batch size, and after 200 epochs for the learning rate experiment. The results in all phase diagrams are averaged over 10 independent runs. Notably, when we replace BatchNorm with the scaling constant $c_\ell$ mentioned in the main text, we still observe similar convex--concave patterns in the phase diagram corresponding to noise and nonlinearity. However, without the adaptive BatchNorm, the load curves exhibit more fluctuations and are less smooth, and the variations between independent runs become more pronounced.

In the left column of FIG.~\ref{fig: deformed-maintext}, we train the described DNN with a learning rate of 0.0001 and the batch size of 200 in panel (b); learning rate of 0.001, dropout of 0.1 and batch size of 100 in panel (c); learning rate of 0.003, and batch size of 50, and dropout of 0.2 in panel (d). In the right column, we set $k=1, y=1$ for all three cases, and $\mu_\rightarrow=0.11,\mu_\leftarrow=0.03$ and $\varepsilon=0.006$ for (b); $\mu_\rightarrow=\mu_\leftarrow=0.04$ and $\varepsilon=0.01$ for (c); $\mu_\rightarrow=0.04,\mu_\leftarrow=0.12$ and $\varepsilon=0.011$ for (d). 
These parameters are chosen to produce qualitatively similar curves with the previous three characteristic training dynamics. For the spring experiments in FIG.~\ref{fig: deformed-maintext}, we use the same noise for all blocks $\xi_\ell(t)=\xi_{\ell'}(t)$, to mimic the training of DNNs in which the randomness (e.g., data noise, learning rate, and batch size) in each layer is not independent. This synchronous noise results in more similar dynamics over epochs (in particular, the fluctuations) but ultimately leads to similar load curves as independent noise, as shown in FIG.~\ref{fig: spring}.

In FIG.~\ref{fig: orth}(a), we adopt the setting from~\cite{yaras2023law}. We use SGD (instead of ADAM) with a learning rate of 0.001 to train a linear DNN. All weight matrices are initialized as random orthogonal matrices; the data is also a random orthogonal matrix with random binary labels. There is no batch normalization. We set the learning rate to $0.01$ to generate the ``large step size'' result (blue curve) and apply $10\%$ dropout to generate the ``dropout'' results (green curve). 

In FIG.~\ref{fig: CNN}, we consider a CNN with 16 channels and 6 convolutional layers on the same MNIST dataset. We use ReLU activations and BatchNorm between the convolutional layers. Pooling and upsampling are applied after each activation in such a way that each intermediate layer has the same shape, and a linear layer is applied after the final convolutional layer. The learning rate is set as 0.0001.  In FIG.~\ref{fig: orth}(c), the CNN has 7 convolutional layers and 20 channels for each layer, and we train the network for 2000 epoches on the CIFAR10 dataset. The other hyperparameters are the same as in the experiments in FIG.~\ref{fig: CNN}.

All experiments are reproducible using code at \url{https://github.com/DaDaCheng/DNN_Spring}.

\section{S4. Some other metrics for data separation of intermediate features}

\begin{figure}[hb!]
    \centering
    \includegraphics[width=0.8\linewidth]{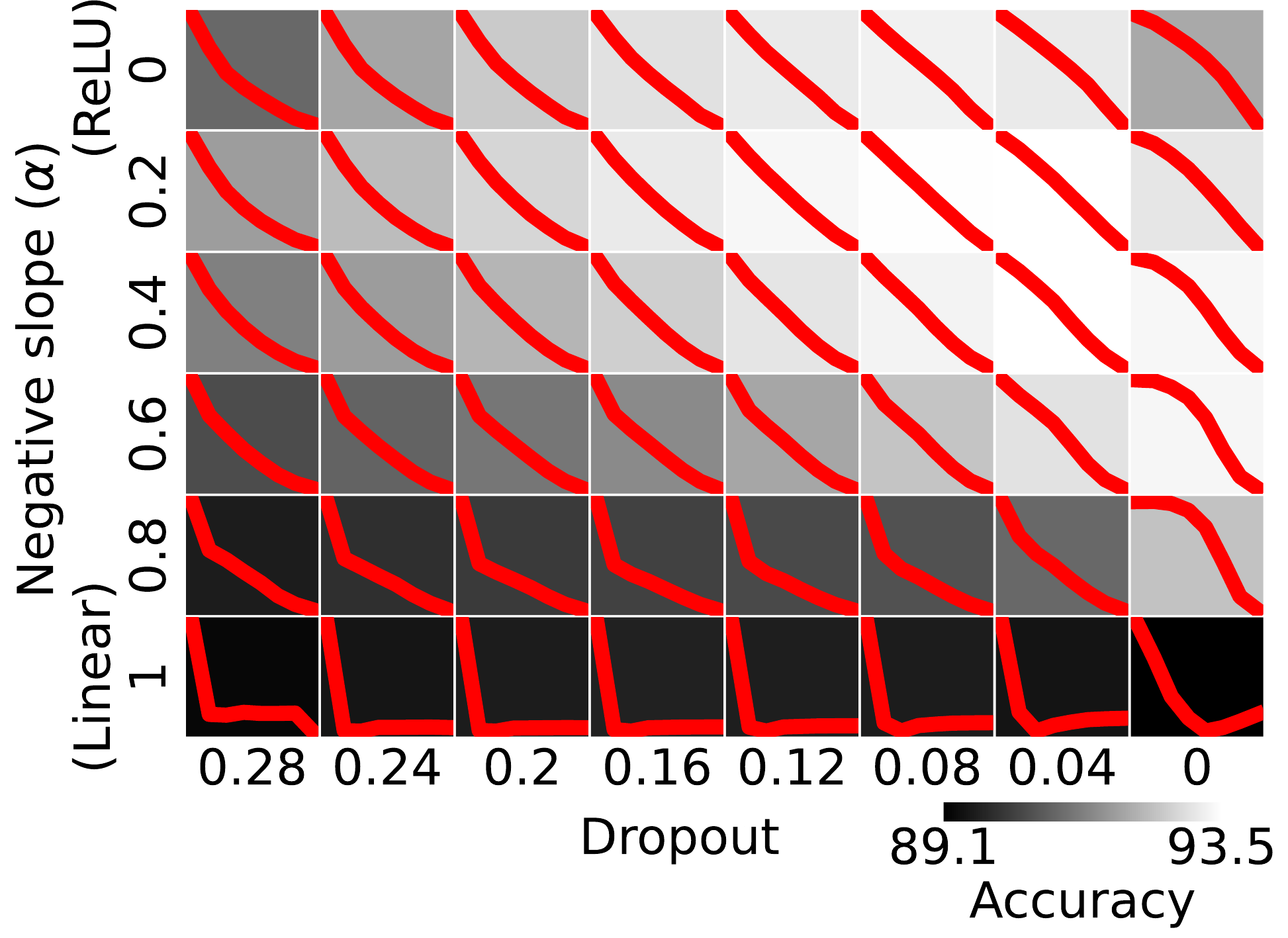}
    \caption{Phase diagram of the DNN load curve on the MNIST for nonlinearity vs. dropout with the data separation as $\hat{D}_\ell$ in Eq.~\eqref{eqn: hat_D}.}
    \label{fig: hat_D}
\end{figure}

\begin{figure}[hb!]
    \centering
    \includegraphics[width=1\linewidth]{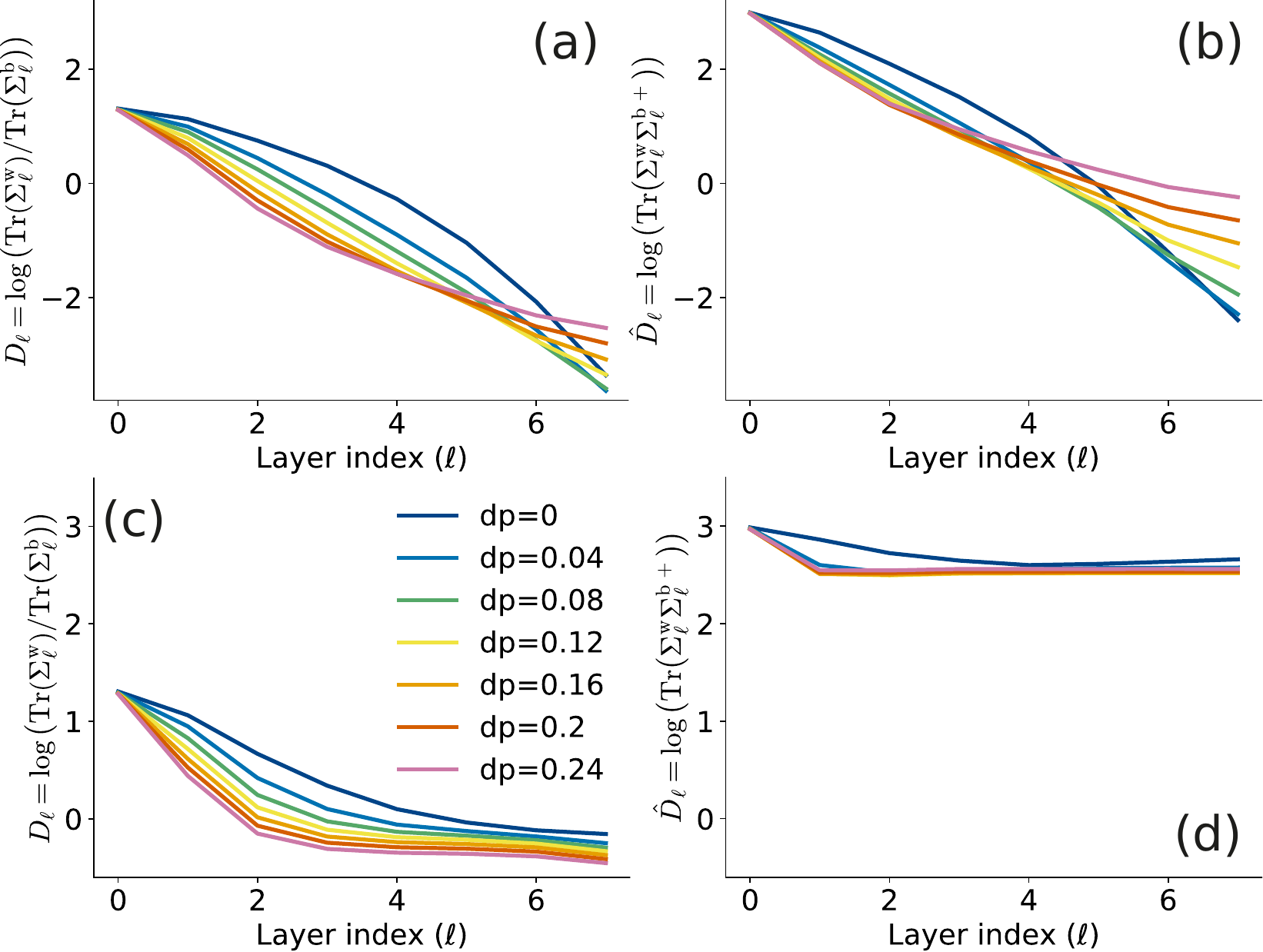}
    \caption{Comparison of data separation metrics $D_{\ell}$ (Eq.~\eqref{eqn: Dell}) in (a, c) and $\hat{D}_{\ell}$ (Eq.~\eqref{eqn: hat_D}) in (b, d). The top two plots (a, b) show results for the ReLU DNN, while the bottom two plots (c, d) show results for the linear DNN. The curves in all four plots represent the same DNNs trained on MNIST with different dropout (dp) rates, as indicated in the legend in (c).}
    \label{fig: diffD}
\end{figure}

In the main text, we use Eq.~\eqref{eqn: Dell} to quantify the ratio between the ``signal'' and the ``noise''. Some works use a different but related quantity ~\cite{he2023law, papyan2020prevalence, zarka2021separation},
\begin{equation}\label{eqn: hat_D}
\hat{D}_{\ell}:= \log\left(\mathrm{Tr}\left(\Sigma_{\ell}^\mathrm{w}{\Sigma_{\ell}^\mathrm{b}}^+\right)\right),
\end{equation}
where $\Sigma^+$ denotes the Moore--Penrose pseudoinverse of $\Sigma$. In general, for our purposes, these two metrics produce comparable results; we show this in FIG.~\ref{fig: hat_D}. 
The behavior of the two metrics shows clear differences in the case of \emph{linear} networks where $\hat{D}_\ell$ cannot capture the differences in features across different layers (FIG.~\ref{fig: diffD} (c) and (d)).



\section{S5. Regression}

\begin{figure}[t!]
    \centering
    \includegraphics[width=1\linewidth]{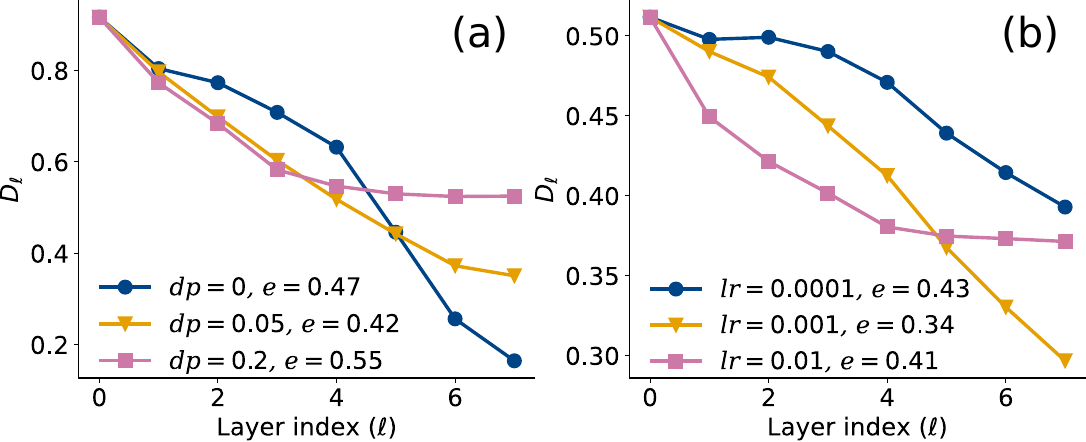}
    \caption{Regression load curves on the diabetes dataset~\cite{efron2004least} \textbf{(a)} and superconductivity critical temperature~\cite{hamidieh2018data} \textbf{(b)} for a 7-hidden-layer ReLU network. In \textbf{(a)}, \textit{dp} denotes the dropout ratio; in \textbf{(b)}, \textit{lr} represents the learning rate; and \textbf{e} indicates the test RMSE.}
    \label{fig: regression}
\end{figure}

\begin{figure}[th!]
    \centering
    \includegraphics[width=1\linewidth]{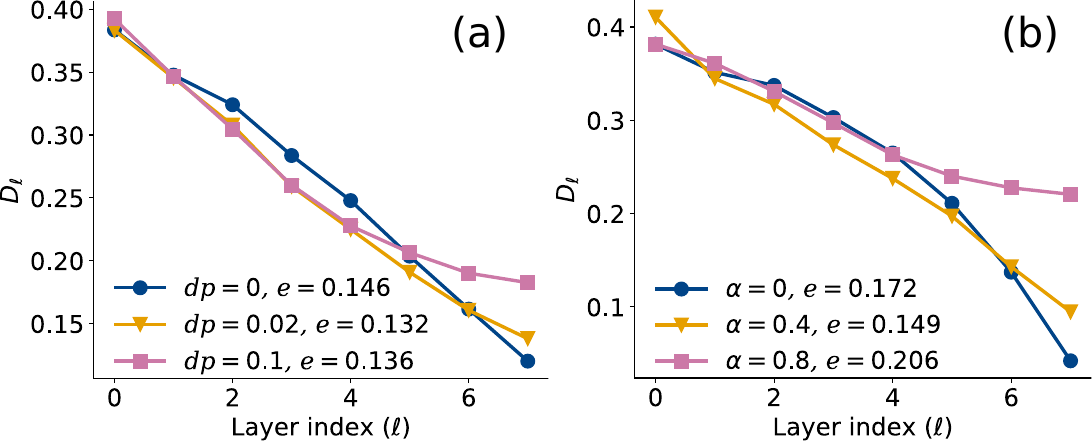}
    \caption{Regression load curves on MNIST dataset trained with one-hot label and MSE loss for a 7-hidden-layer ReLU/LeakyReLU network. In \textbf{(a)}, \textit{dp} denotes the dropout ratio and MLPs have ReLU activation function; in \textbf{(b)}, \textbf{$\alpha$} represents the native slope of LeakyReLU; and \textbf{e} indicates the test RMSE.}
    \label{fig: regression_MNIST}
\end{figure}

The phenomenology described in the main text for classification can be observed for regression problems if we define the load as the MSE (or RMSE) of the optimal linear regressor from the $\ell$th layer features,
\begin{equation*}
   D_\ell := \sqrt{\min_{\vw,\vb} \frac{1}{2N} \sum_{i=1}^N \Vert \vy_i-\vw^\top(\vx_i)_{\ell}+\vb\Vert^2_2},
\end{equation*}
where $N$ is the number of data pairs and the summation is taken over all data. As shown in FIG.~\ref{fig: regression}, noisier training results in a convex load curve, whereas lazy training results in a concave curve. Furthermore, the straighter line shows better generalization.
We also test the MNIST dataset in a regression setting by using one-hot labels. The regression load curves consistently align with the phenomenon observed in the classification task in FIG.~\ref{fig: regression_MNIST}.

\section{S6. Additional details about the spring block systems}
\subsection{Friction in the second order system}

\begin{figure}[t]
    \centering
    \includegraphics[width=1\linewidth]{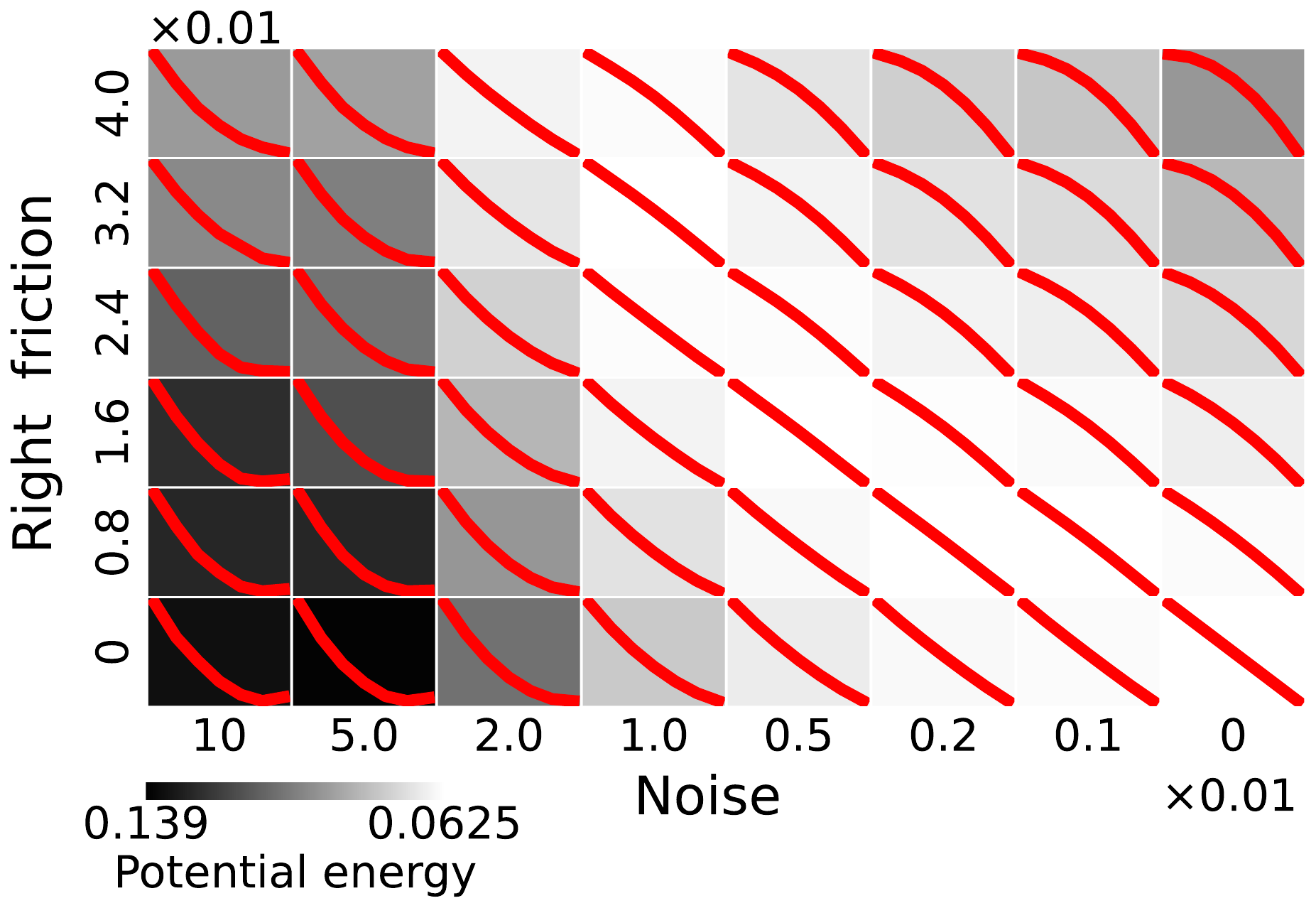}
    \caption{Phase diagram of the second order spring block system \eqref{eqn: second order dynamics} for friction $\mu_\rightarrow$ vs. noise level $\epsilon$. We set $k=\gamma=1,\mu_\leftarrow=0.12$ and $L=7$. Red load curves  ($D_\ell=y-x_\ell$) are recorded at $t=100$; the shading corresponds to the elastic potential energy.}
    \label{fig: spring phase diagram second}
\end{figure}

To build intuition about the second order system dynamics~\eqref{eqn: second order dynamics}, we can rewrite it as
\begin{equation}
\label{eqn: deterministic dynamics SM}
\begin{aligned}
    F_\ell  &= k(x_{\ell+1}-2x_\ell+x_{\ell-1}) -\gamma v_{\ell}+ \varepsilon \xi_\ell,\\
    \ddot{x}_\ell &= F_\ell +f_\ell=\sigma(F_\ell; \dot{x}_\ell)\\
\end{aligned}
\end{equation}
where the friction $f_\ell$ depends on the force $F_\ell$ and the velocity $\dot{x}_\ell$. If a block is moving to the right ($\dot{x}_\ell > 0$), sliding friction resists its movement as $f_\ell = -\mu_\rightarrow$. If a block is moving to the left ($\dot{x}_\ell < 0$), sliding friction is $f_\ell = \mu_\leftarrow$. When a block is stationary ($\dot{x} = 0$), static friction compensates for all other forces as long as they do not exceed the maximum static friction, i.e., $f_\ell = -F_\ell$ when $-\mu_\leftarrow \leq F_\ell \leq \mu_\rightarrow$. We take the maximum static friction to be equal to the sliding friction to simplify the model. We can summarize the above cases in an activation-like form,
\begin{equation}
\begin{aligned}
     \sigma(z;\dot{x})&=\begin{cases}
           0 & \text{~if~} \dot{x}=0 \text{~and~}  -\mu_{\leftarrow}\leq z \leq \mu_{\rightarrow}\\
           z-\mu_{\rightarrow} & \text{~if~} \dot{x}> 0 \text{~or~} (\dot{x}= 0 \text{~and~} z > \mu_\rightarrow)\\
           z+\mu_{\leftarrow} & \text{~if~}  \dot{x}< 0 \text{~or~} (\dot{x}= 0 \text{~and~} z <-\mu_{\leftarrow}).\\
           \end{cases}
           \end{aligned}
\end{equation}
The phase diagram of this second order system is shown in FIG.~\ref{fig: spring phase diagram second}.

\subsection{Noisy equilibiria and separation of time scales}

In the main text, we mentioned that it is convenient to assume bounded (zero-mean, symmetric) noise increments, for example by using a truncated Gaussian distribution. Without truncation, the trajectories of the spring--block system exhibit two stages. In the first stage, the elastic force dominates the Gaussian tails and the block motion is primarily driven by the (noisy) spring force. At the end of this period, the spring force is balanced with friction. At this point the blocks can only move due to a large realization of noise. These low-probability realizations will move the blocks very slowly close to the equilibrium $\Delta d^* = \frac{\mu_\rightarrow - \mu_\rightarrow}{2k}$ which is stable under symmetric noise perturbations. This is undesirable for analysis since this stable point does not depend on the noise level; in particular, it will be eventually reached even for arbitrarily small $\varepsilon$. This, however, will happen in an exponentially long time, longer than $e^{c_0/\epsilon^2}$ for some constant $c_0$. We can obviate this nuisance in three natural ways: by assuming bounded noise incremenents, by applying noise decay, or by introducing a stopping criterion, for example via a relative change threshold. All are well-rooted in DNN practice: common noise sources are all bounded, and standard training practices involve a variety of parameter scheduling and stopping criteria. 

\subsection{Second order Langevin equation}

\begin{figure}[h!]
    \centering
    \includegraphics[width=1\linewidth]{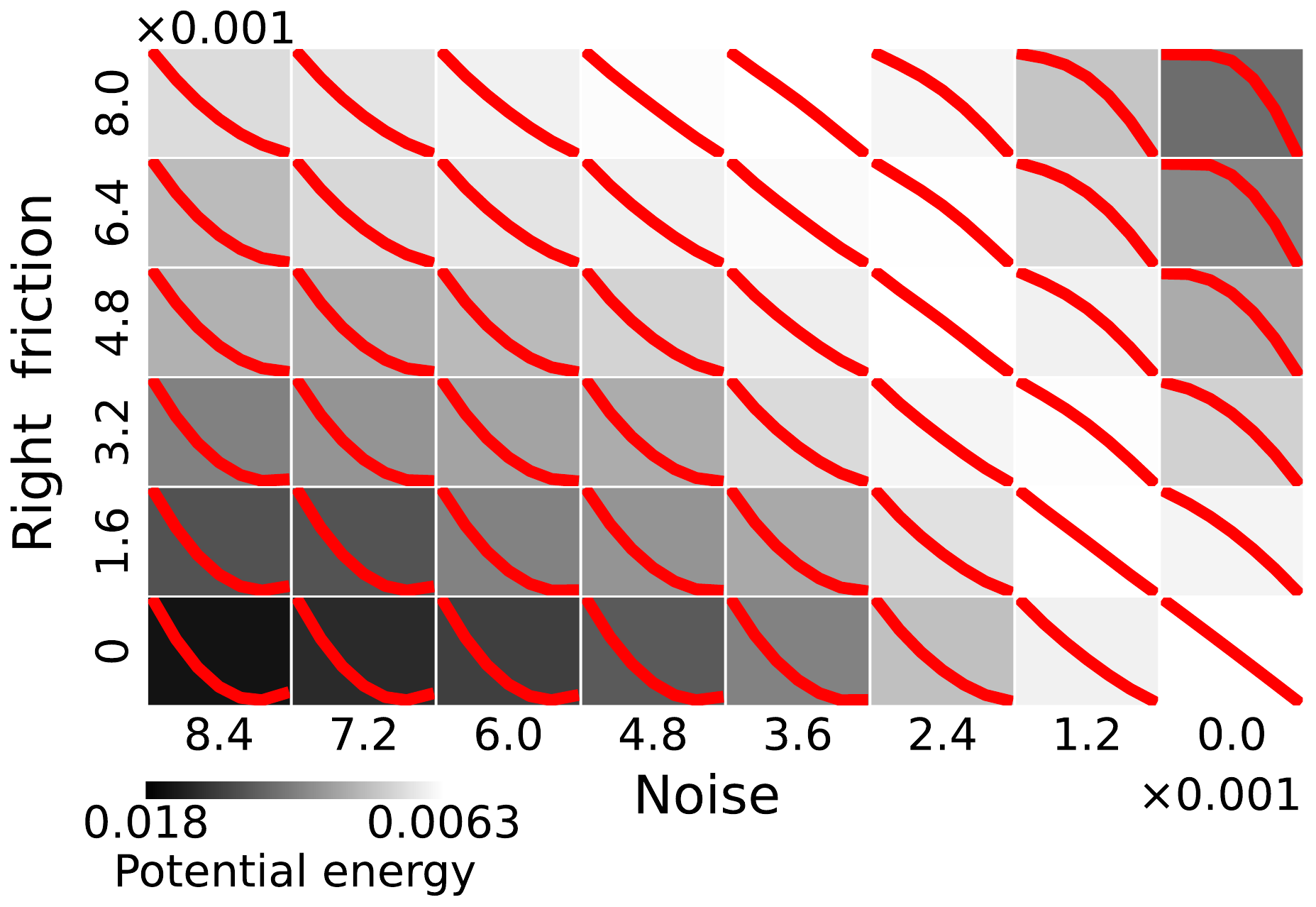}
    \caption{Phase diagram of spring block system with Langevin equation \eqref{eqn: SDE} for friction $\mu_\rightarrow$ vs. noise level $\epsilon_0$. We set $k=0.1, \gamma=1, \tau=0.002,\mu_\leftarrow=0.016$ and $L=7$. Red load curves  ($D_\ell=y-x_\ell$) are recorded at $t=300$; the shading corresponds to the elastic potential energy.} 
    \label{fig: SDE}
\end{figure}

We can obtain a more standard Langevin dynamics formulation of Eq.~\eqref{eqn: deterministic dynamics SM} by adding noise to the velocity independently of the friction, i.e., computing the friction before adding noise. We note that this is less realistic from a physical point of view. The position of the $\ell$th block $x_{\ell}=\sum_{i=1}^\ell d_{i}$ then obeys the equation of motion $\mathrm{d}x=v \mathrm{d}t$ with
\begin{equation}\label{eqn: SDE}
    \mathrm{d} v_\ell = \left(k(\boldsymbol{L}x)_\ell -f(v_\ell) - \gamma v_\ell \right) \mathrm{d}t + \varepsilon(t) \mathrm{d} W_\ell(t),
\end{equation}
where the sliding friction $f$ resists movement as
\begin{equation}\label{eqn: Langevin Friction}
    f(v)=\begin{cases}
        \mu_\rightarrow &\text{if~} v>0\\
        -\mu_\leftarrow &\text{if~} v<0.
    \end{cases}
\end{equation}
In Eq. \eqref{eqn: SDE},  $W_\ell(t)$ represents the Wiener process with $\varepsilon$ controlling the amount of the noise (temperature parameters). To ensure convergence in this formulation we have to decay the noise; we set $\varepsilon(t)=\varepsilon_0 e^{-\tau t}$. The friction at zero speed $f(0)$ is defined similarly to the static friction $f_\ell$ in Eq.~\eqref{eqn: deterministic dynamics SM}, but with $F_\ell=k(\mL x)_\ell$ without considering noise. This dynamics can be solved by standard SDE integration. It results in a similar phase diagram at convergence, as shown in FIG.~\ref{fig: SDE}, although the fluctuations do not appear as similar as with the formulation in the main text.

\section{S7. Scaling and quantity of noise sources}
In FIG.~\ref{fig: pd}, we show that four ``noise'' sources—data noise, learning rate, dropout, and batch size—exhibit similar phase diagrams. Although it is difficult to precisely quantify these different factors in deep neural networks in an integrated manner, their scaling (whether logarithmic, linear, or otherwise on the x-axis of the phase diagram) can still be verified using known results on stochastic modified equations~\cite{li2017stochastic}. Specifically,  since SGD can be approximated (in a certain weak sense) by a stochastic differential equation (SDE),  by analyzing the structure of the drift and diffusion terms in this SDE, we can understand the role and scaling of different types of noise. Here, we use the learning rate as a reference to examine how other hyperparameters should be adjusted when varying the learning rate to keep the SDE unchanged. For example, standard SGD can be written as
\begin{equation*} \vtheta_{k+1}=\vtheta_{k} -\eta \nabla_{\vtheta}\left(\frac{1}{|\mathcal{B}_k|} \sum_{i\in \mathcal{B}_k} l\left(\vtheta_k; (\vx_i,\vy_i)\right)\right), \end{equation*}
where $\eta$ is the learning rate, and the mini-batch $\mathcal{B}_k$ is sampled uniformly from a large dataset. For a typical loss function, neural network, and data distribution $(\vx_i,\vy_i)$, we can treat $\nabla{\vtheta} l(\vtheta; (\vx_i,\vy_i) )$ as a random variable with mean $\vg(\vtheta)$ and covariance $\mC(\vtheta)$. For a large batch size $B=|\mathcal{B}_k|$, by the central limit theorem the random fluctuations are approximately normal $\vxi^C_k\sim\mathcal{N}(0,\mC(\vtheta_k))$, so that the SGD steps (approximately) read
\begin{equation*} \vtheta_{k+1}=\vtheta_{k} -\eta \vg(\vtheta_k) + \frac{\eta}{\sqrt{B}} \vxi^C_k. \end{equation*}
This can be recognized as the Euler--Maruyama discretization of the following SDE
\begin{equation*} \mathrm{d} \vtheta=-\vg(\vtheta)\mathrm{d}t + \sqrt{\frac{\eta}{B}} \mC^{\frac{1}{2}}(\vtheta)\mathrm{d}\mW(t), \end{equation*}
where $\mW(t)$ is a standard Brownian motion. In this case, if we plot the learning rate $\eta$ on a logarithmic scale, the batch size $B$ should also be plotted on a logarithmic scale. This relationship has been observed and studied in earlier works, e.g.~\cite{goyal2017accurate,smith2017bayesian,jastrzkebski2017three}.

Dropout in deep neural networks (DNNs) is more complex, as the stochasticity occurs at each layer and accumulates across the depth. A recent work~\cite{zhang2024stochastic} analyzed this problem in a two-layer network with MSE loss. The authors approximate the trajectories of full batch GD with dropout by the following SDE:
\begin{equation*}
\begin{aligned}
\mathrm{d} \vtheta &= -\left( \vg_1(\vtheta)+ \frac{q}{1-q} \vg_2(\vtheta) \right) \mathrm{d}t\\ &+ \sqrt{\eta} \left( \frac{q}{1-q}\mC_1(\vtheta )+ \frac{q}{(1-q)^2} \mC_2(\vtheta) \right)^{1/2} \mathrm{d}\mW(t),
\end{aligned}
\end{equation*}
where $q$ is the probability of setting activations to zero. For small dropout, as $q \to 0$, we can approximate it as
$
\mathrm{d} \vtheta\approx -\vg_1(\vtheta)\mathrm{d}t + \sqrt{\eta q}(\mC_1+\mC_2)^{\frac{1}{2}} \mathrm{d}\mW(t).
$
It implies that if we plot the learning rate $\eta$ on a logarithmic scale, the dropout ratio $q$ should also be plotted on a logarithmic scale.

In the data noise experiments, we randomly reset a fraction $p$ of the labels while simultaneously adding Gaussian noise with variance $p^2$ to the input data. When $p$ is away from zero, the noise introduced by label corruption is significantly larger than the noise from the input data. Therefore, we focus primarily on analyzing label noise. Since cross-entropy loss is biased and scale-sensitive~\cite{maennel2020neural}, most analytical studies on label noise are based on regression settings with unbiased Gaussian noise, e.g.,~\cite{chen2023stochastic,damian2021label}. However, we can still analyze a simple setting to estimate the scale based on the method we used previously. 

To simplify the problem, we study a single data point $\vx$ with label $1$ and a one-layer network $f_m(x) = \langle \vw_m, \vx \rangle$ with $m = 1, \dots, M$, corresponding to $M$ classes.  The cross-entropy loss with softmax can be computed as 
\begin{equation*}
   \mathcal{L}=\begin{cases}  
   -\log \frac{\exp \langle \vw_1, \vx \rangle}{\sum_m \exp \langle \vw_m, \vx \rangle}  &\quad \text{w. p. $1 - p + \frac{p}{M}$,}\\
     -\log \frac{\exp \langle \vw_j, \vx \rangle}{\sum_m \exp \langle \vw_m, \vx \rangle} &\quad \text{for $j > 1$, w. p. $\frac{p}{M}$.}
    \end{cases}
\end{equation*}
We can then compute the mean and variance of each gradient descent step and approximate the GD by the stochastic modified equations SDE as mentioned in~\cite{li2017stochastic,zhang2024stochastic}. After rescaling the loss function, and assuming the number of classes is large, $M \gg 1$, we obtain a simple formula:
\begin{equation*}
    \mathrm{d}\vw_1= \left(1 - \frac{1}{1 - p} t_1(\vw)\right)\vx + \sqrt{\eta \frac{p}{1 - p}} \left(\vx\vx^T\right)^{\frac{1}{2}} \mathrm{d}\mW(t),
\end{equation*}
where $t_1 = \frac{\exp \langle \vw_1, \vx \rangle}{\sum_m \exp \langle \vw_m, \vx \rangle}$. The stochasticity of the remaining weights $\vw_m$ for $m > 1$ is negligible for large $M$. We also note that $t_1$ is small at initialization, so to obtain a consistent phase diagram, it is better to set $\frac{p}{1 - p}$ in logarithmic scale when using $\eta$ in logarithmic scale. Noting that we are not only interested in small $p$, we can leverage the well-known approximation of the logit function:
\begin{equation*}
\mathrm{logit}(x) = \log\left(\frac{x}{1 - x}\right) \approx 4(x - 0.5) \quad \text{for } x \text{ near } 0.5,
\end{equation*}
which leads to a linear scale of x-axis in the first panel in FIG.~\ref{fig: pd}.

\end{document}